\documentclass[twocolumn,showpacs,aps,prd]{revtex4-1}

\RequirePackage[colorlinks,breaklinks,pdftitle={ATLAS draft},pdfauthor={The ATLAS Collaboration}]{hyperref}  
\hypersetup{linkcolor=blue,citecolor=blue,filecolor=black,urlcolor=blue}

\usepackage{graphicx} 
\graphicspath{{figures/}}
\usepackage{atlasphysics} 
\usepackage{rotating} 
\usepackage{multirow}

\newcommand{\Rhad}    {\ensuremath{R_{\rm had}}}

\newcommand{\Reta}    {\ensuremath{R_{\eta}}}
\newcommand{\Rphi}    {\ensuremath{R_{\phi}}}
\newcommand{\wetatwo} {\ensuremath{w_{2}}}

\newcommand{\Fside}   {\ensuremath{F_{\rm side}}}
\newcommand{\wthree}  {\ensuremath{w_{s,3}}}
\newcommand{\wtot}    {\ensuremath{w_{s,{\rm tot}}}}
\newcommand{\DeltaE}  {\ensuremath{\Delta{}E}}
\newcommand{\Eratio}  {\ensuremath{E_{\rm ratio}}}

\newcommand{\Etiso}   {\ensuremath{E_{\mathrm{T}}^{\mathrm{iso}}}}

\newcommand{\mcc}[1]{\multicolumn{1}{c}{#1}}

\begin{document}

%
%
\begin{minipage}[t]{\textwidth}
  \begin{flushright}
    CERN-PH-EP-2010-068\\
    Submitted to Phys. Rev. D
    \end{flushright}
\end{minipage}

\vspace{0.5cm}

\title{Measurement of the inclusive isolated prompt photon cross section in $pp$ collisions at $\sqrt{s}= 7\TeV$ with the ATLAS detector}

\author{G. Aad \textit{et al.}}\thanks{Full author list given at the end of the article in Appendix~\ref{app:ATLASColl}.}
\collaboration{The ATLAS Collaboration}

\date{\today}

\begin{abstract}
  A measurement of the cross section for the
  inclusive production of isolated prompt photons in $pp$ collisions
  at a centre-of-mass energy $\sqrt{s} = 7\TeV$ is presented.
  The measurement covers the pseudorapidity ranges $|\eta^\gamma|<1.37$ 
  and $1.52\leq|\eta^\gamma|<1.81$ in 
  the transverse energy range $15\leq E_{\rm T}^{\gamma}<100 \GeV$. The results
  are based on an integrated luminosity of 880~\inb, collected with
  the ATLAS detector at the Large Hadron Collider.
  Photon candidates are identified by combining information from the
  calorimeters and from the inner tracker.
  Residual background in the selected sample is estimated from data
  based on the observed distribution
  of the transverse isolation energy in a narrow cone around the photon candidate.
  The results are compared to predictions from next-to-leading
  order perturbative QCD calculations.
\end{abstract}

\pacs{13.25.Hw, 12.15.Hh, 11.30.Er}

\maketitle


\newpage

\section{Introduction}
\label{sec:Introduction}
%
%

Prompt photon production at hadron colliders provides a handle for
testing perturbative QCD (pQCD)
predictions~\cite{Angelis_promptphoton,QCD_promptphoton_NLO}.  Photons
provide a colorless probe of quarks in the hard partonic interaction
and the subsequent parton shower. Their production is directly
sensitive to the gluon content of the proton through the
$qg~\rightarrow~q\gamma$ process, which dominates at leading order
(LO).  The measurement of the prompt photon production cross section
can thus be exploited to constrain the gluon density
function~\cite{Akesson_promptphoton,promptphoton_and_gluon_pdf}.
Furthermore, photon identification is important for many physics
signatures, including searches for Higgs
boson~\cite{Higgs:1964ia,Englert:1964et,Guralnik:1964eu}, graviton
decays~\cite{Randall:1999ee} to photon pairs, decays of excited
fermions~\cite{Weinberg:1979bn}, and decays of pairs of supersymmetric
particles characterized by the production of two energetic photons and
large missing transverse
energy~\cite{Dine:1981za,Dimopoulos:1981au,Nappi:1982hm}.

%
%

Prompt photons include both ``direct'' photons, which take part in the
hard scattering subprocess (mostly quark-gluon Compton scattering,
$qg\to q\gamma$, or quark-antiquark annihilation, $q\bar{q} \to
g\gamma$), and ``fragmentation'' photons, which are the result of the
fragmentation of a high-$\pt$ parton~\cite{jetphox1,jetphox2}.  In
this analysis, an isolation criterion is applied based on the amount
of transverse energy inside a cone of radius $R = \sqrt{\left( \eta -
\eta^{\gamma} \right)^{2} + \left( \phi - \phi^{\gamma} \right)^2} =
0.4$ centered around the photon direction in the pseudorapidity
($\eta$) and azimuthal angle ($\phi$) plane~\footnote{The ATLAS
reference system is a Cartesian right-handed coordinate system, with
the nominal collision point at the origin. The anticlockwise beam
direction defines the positive $z$-axis, while the positive $x$-axis
is defined as pointing from the collision point to the centre of the
LHC ring and the positive $y$-axis points upwards. The azimuthal angle
$\phi$ is measured around the beam axis, and the polar angle $\theta$
is measured with respect to the $z$-axis. Pseudorapidity is defined as
$\eta = -\ln\tan(\theta/2)$.}.  After the isolation requirement is
applied the relative contribution to the total cross section from
fragmentation photons decreases, though it remains non-negligible
especially at low transverse energies~\cite{jetphox2}.  The isolation
requirement also significantly reduces the main background of
non-prompt photon candidates from decays of energetic $\pi^{0}$ and
$\eta$ mesons inside jets.

%
%

Early studies of prompt photon production were carried out at the ISR
collider~\cite{Anassontzis:1982gm,Angelis:1989zv}.  Subsequent
studies, for example
\cite{Appel:1986ix,UA1_promptphoton,Werlen:1999ij}, further
established prompt photons as a useful probe of parton interactions.
More recent measurements at hadron colliders were performed at the
Tevatron, in $p\bar{p}$ collisions at a centre-of-mass energy
$\sqrt{s}=1.96$ TeV.  The measurement by the D0
Collaboration~\cite{D0_promptphoton} is based on 326~\ipb~and covers a
pseudorapidity range $|\eta^\gamma|<0.9$ and a transverse energy range
$23<\ET^\gamma<300~\GeV$, while the measurement by the CDF
Collaboration~\cite{CDF_promptphoton} is based on 2.5~\ifb~and covers
a pseudorapidity range $|\eta^\gamma|<1.0$ and a transverse energy
range $30<\ET^\gamma<400~\GeV$.  Both D0 and CDF measure an isolated
prompt photon cross section in agreement with next-to-leading order
(NLO) pQCD calculations, with a slight excess seen in the CDF data
between 30 and 50 GeV.  Measurements of the inclusive prompt photon
production cross section have also been performed in $ep$ collisions,
both in photoproduction and deep inelastic scattering, by the
H1~\cite{H1_promptphoton,HoneDIS} and
ZEUS~\cite{ZEUSPhotoprod,ZEUS_promptphoton} Collaborations.  The most
recent measurement of the inclusive isolated prompt photon production
was done with 2.9~\ipb~at $\sqrt{s}=7$ TeV by the CMS
Collaboration~\cite{CMS_prompt_photons}.  That measurement, which
covers $21<\ET^\gamma<300~\GeV$ and $|\eta^\gamma|<1.45$, is in good
agreement with NLO predictions for the full $\ET^{\gamma}$ range.

%
%

This paper describes the extraction of a signal of isolated prompt
photons using 880~\inb~of data collected with the ATLAS detector at
the Large Hadron Collider (LHC).  A measurement of the production
cross section in $pp$ collisions at $\sqrt{s} = 7~\TeV$ is presented,
in the pseudorapidity ranges $|\eta^\gamma|<0.6$, $0.6\leq
|\eta^\gamma|<1.37$ and $1.52\leq |\eta^\gamma| < 1.81$, for photons
with transverse energies between 15~\GeV~and 100~\GeV.

%
%

The paper is organized as follows. The detector is described in
Section~\ref{sec:Detector}, followed by a summary of the data and the
simulated samples used in the measurement in
Section~\ref{sec:Samples}.  Section~\ref{sec:Theory} introduces the
theoretical predictions to which the measurement is compared.
Section~\ref{sec:PhotonID} describes the photon reconstruction and
identification algorithms; their performance is given in
Section~\ref{sec:Efficiency}. Section~\ref{sec:Purity} describes the
methods used to estimate the residual background in the data and to
extract the prompt photon signal, followed by a discussion of the data
corrections for the cross section measurement in
Section~\ref{sec:Unfolding}. The sources of systematic uncertainties
on the cross section measurement are discussed in
Section~\ref{sec:Systematics}. Section~\ref{sec:CrossSection} contains
the main experimental results and the comparison of the observed cross
sections with the theoretical predictions, followed by the conclusions
in Section~\ref{sec:Conclusion}.

\section{The ATLAS detector}
\label{sec:Detector}
The ATLAS detector is described in detail in
Refs.~\cite{ATLAS_detector} and~\cite{ATLAS_CSC}.  For the measurement
presented in this paper, the calorimeter, with mainly its
electromagnetic section, and the inner detector are of particular
relevance.

The inner detector consists of three subsystems: at small radial
distance $r$ from the beam axis ($50.5<r<150$ mm), pixel silicon
detectors are arranged in three cylindrical layers in the barrel and
in three disks in each end-cap; at intermediate radii ($299<r<560$
mm), double layers of single-sided silicon microstrip detectors are
used, organized in four cylindrical layers in the barrel and nine
disks in each end-cap; at larger radii ($563<r<1066$ mm), a straw
tracker with transition radiation detection capabilities divided into
one barrel section (with 73 layers of straws parallel to the beam
line) and two end-caps (with 160 layers each of straws radial to the
beam line) is used.  These three systems are immersed in a 2 T axial
magnetic field provided by a superconducting solenoid.  The inner
detector has full coverage in $\phi$. The silicon pixel and microstrip
subsystems cover the pseudorapidity range $|\eta|<2.5$, while the
transition radiation tracker (TRT) acceptance is limited to the range
$|\eta|<2.0$.  The inner detector allows an accurate reconstruction of
tracks from the primary proton-proton collision region, and also
identifies tracks from secondary vertices, permitting the efficient
reconstruction of photon conversions in the inner detector up to a
radius of $\approx$~80~cm.

The electromagnetic calorimeter is a lead-liquid argon (Pb-LAr)
sampling calorimeter with an accordion geometry.  It is divided into a
barrel section, covering the pseudorapidity region $|\eta|< 1.475$,
and two end-cap sections, covering the pseudorapidity regions
$1.375<|\eta|<3.2$.  It consists of three longitudinal layers.  The
first one, with a thickness between 3 and 5 radiation lengths, is
segmented into high granularity strips in the $\eta$ direction (width
between 0.003 and 0.006 depending on $\eta$, with the exception of the
regions $1.4<|\eta|<1.5$ and $|\eta|>2.4$), sufficient to provide an
event-by-event discrimination between single photon showers and two
overlapping showers coming from a $\pi^0$ decay.  The second layer of
the electromagnetic calorimeter, which collects most of the energy
deposited in the calorimeter by the photon shower, has a thickness
around 17 radiation lengths and a granularity of $0.025\times0.025$ in
$\eta\times\phi$ (corresponding to one cell).  A third layer, with
thickness varying between 4 and 15 radiation lengths, is used to
correct leakage beyond the calorimeter for high energy showers. In
front of the accordion calorimeter a thin presampler layer, covering
the pseudorapidity interval $|\eta|<1.8$, is used to correct for
energy loss before the calorimeter. The sampling term $a$ of the
energy resolution ($\sigma(E)/E$ $\approx$ $a/\sqrt{E~{\rm [GeV]}}$)
varies between 10\% and 17\% as a function of $|\eta|$ and is the
largest contribution to the resolution up to about 200 GeV, where the
global constant term, estimated to be 0.7\%~\cite{Winter_confNote},
starts to dominate.

The total amount of material before the first active layer of the
electromagnetic calorimeter (including the presampler) varies between
2.5 and 6 radiation lengths as a function of pseudorapidity, excluding 
the transition region ($1.37\leq|\eta|<1.52$) between the barrel and the
end-caps, where the material thickness increases to 11.5 radiation lengths.
The central region ($|\eta|<0.6$) has significantly less material than
the outer barrel ($0.6\leq|\eta|<1.37$), which motivates the division
of the barrel into two separate regions in pseudorapidity.

%
%

A hadronic sampling calorimeter is located beyond the electromagnetic
calorimeter.  It is made of steel and scintillating tiles in the
barrel section ($|\eta|<1.7$), with depth around 7.4 interaction
lengths, and of two wheels of copper and liquid argon in each end-cap,
with depth around 9 interaction lengths.

%
%

A three-level trigger system is used to select events containing
photon candidates during data taking.  The first level trigger
(level-1) is hardware based: using a coarser cell granularity
($0.1\times 0.1$ in $\eta\times\phi$) than that of the electromagnetic
calorimeter, it searches for electromagnetic clusters within a fixed
window of size $0.2\times 0.2$ and retains only those whose total
transverse energy in two adjacent cells is above a programmable
threshold.  The second and third level triggers (collectively referred
to as the ``high-level'' trigger) are implemented in software.  The
high-level trigger exploits the full granularity and precision of the
calorimeter to refine the level-1 trigger selection, based on improved
energy resolution and detailed information on energy deposition in the
calorimeter cells.

\section{Collision Data and Simulated Samples}
\label{sec:Samples}
%
%
\subsection{Collision Data}
%
%
The measurement presented here is based on proton-proton collision
data collected at a centre-of-mass energy $\sqrt{s}=7$ TeV 
between April and August 2010.
Events in which the calorimeters or the inner detector are not fully 
operational, or show data quality problems, are excluded.
%
%
Events are triggered using a single-photon high-level trigger with
a nominal transverse energy threshold of 10 GeV, seeded by a level-1 
trigger with nominal threshold equal to 5 GeV.
The selection criteria applied by the trigger on shower shape variables
computed from the energy profiles of the showers in the calorimeters 
are looser than the photon identification 
criteria applied in the offline analysis, and allow ATLAS
to reach a plateau of constant efficiency close to 100\%
for true prompt photons with $\ET^\gamma>15$ GeV and pseudorapidity
$|\eta^\gamma|<1.81$.
In addition, samples of minimum-bias events, triggered by using two 
sets of scintillator counters located at $z=\pm3.5$~m from the 
collision centre, are used to estimate the single-photon trigger efficiency.
%
%
The total integrated luminosity of the sample passing data quality and
trigger requirements amounts to $(880\pm 100)$~\inb.

%
%
In order to reduce non-collision backgrounds, events are required to
have at least one reconstructed primary vertex consistent with the average beam
spot position and with at least three associated tracks.  
The efficiency of this requirement is expected to be greater than
99.9\% in events containing a prompt photon with $\ET^\gamma>15$ GeV and lying within the calorimeter acceptance.
The total number of selected events in data after this requirement is
9.6 million.
%
%
The remaining amount of non-collision background
is estimated using control samples collected with dedicated, 
low threshold triggers that are activated
in events where either no proton bunch or only one of the two beams crosses
the interaction region. The estimated contribution to the
final photon sample is less than 0.1\% and is therefore neglected.

%
%
\subsection{Simulated events}
\label{subsec:simulated_data}
%
%
To study the characteristics of signal and background events,
Monte Carlo (MC) samples are generated using {\tt PYTHIA} 6.4.21~\cite{pythia}, 
a leading-order parton-shower MC generator, with the modified leading
order MRST2007~\cite{MRST2007} parton distribution functions
(PDFs). It accounts for QED radiation emitted off quarks in the initial state
(ISR) and in the final state (FSR).
{\tt PYTHIA} simulates the underlying event using the
multiple-parton interaction model, and uses the Lund string model for
hadronisation~\cite{LundStrings}. 
The event generator parameters are set according to the ATLAS MC09
tune~\cite{ATLAS_MC09_tune}, and the detector response is
simulated using the {\tt GEANT4} program~\cite{geant}.
These samples are then reconstructed with the same algorithms
used for data. More details on the event generation and simulation 
infrastructure are provided in
Ref.~\cite{ATLAS_simulation}. 
For the study of systematic uncertainties related to the choice of the event
generator and the parton shower model, alternative samples are
also generated with {\tt HERWIG} 6.5~\cite{Herwig}. This generator also 
uses LO pQCD matrix elements, but has a different parton shower model 
(angle-ordered instead of \pt-ordered), a different hadronisation model 
(the cluster model) and a different 
underlying event model, which is generated using the {\tt JIMMY} 
package~\cite{JIMMY} with multiple parton interactions enabled.
The {\tt HERWIG} event generation parameters are also set 
according to the MC09 tune.

To study the main background processes, simulated samples of
all relevant 2$\rightarrow$2 QCD hard subprocesses involving only
partons are used.
The prompt photon contribution arising from initial and final state
radiation emitted off quarks is removed from these samples
in studies of the background.

Two different simulated samples are employed to study the properties
of the prompt photon signal.  The first sample consists of
leading-order $\gamma$-jet events, and contains primarily direct
photons produced in the hard subprocesses $q g \rightarrow q \gamma$
and $q \bar{q} \rightarrow g \gamma$.  The second signal sample
includes ISR and FSR photons emitted off quarks in all 2$\rightarrow$2
QCD processes involving only quarks and gluons in the hard scatter.
This sample is used to study the contribution to the prompt photon
signal by photons from fragmentation, or from radiative corrections to
the direct process, that are less isolated than those from the LO
direct processes.

Finally, a sample of $W\rightarrow e\nu$ simulated events
is used for the efficiency and purity studies involving electrons
from $W$ decays.

\section{Theoretical Predictions}
\label{sec:Theory}
The expected isolated prompt photon production cross section as a
function of the photon transverse energy $\ET^\gamma$ is
calculated with the {\tt JETPHOX} Monte Carlo
program~\cite{jetphox1}, which implements a full NLO QCD calculation
of both the direct and the fragmentation contributions to the total cross
section.
In the calculation performed for this measurement,
the total transverse energy carried by the partons inside a cone of
radius $R=0.4$ in the $\eta-\phi$ space
around the photon direction is required to be less than 4~\GeV.
The NLO photon fragmentation function~\cite{Bourhis:2000gs} and the
CTEQ 6.6 parton density functions~\cite{cteq} provided by the {\tt
LHAPDF} package~\cite{LHAPDF} are used.
The nominal renormalization ($\mu_R$), factorization ($\mu_F$) and
fragmentation ($\mu_f$) scales are set to the photon transverse
energy $\ET^\gamma$. 
Varying the CTEQ PDFs within the 68\% C.L. intervals causes
the cross section to vary between 5\% and 2\% as \ET~increases
between 15 and 100~\GeV.
The  variation of the three scales independently between 0.5
and 2.0 times the nominal scale changes the predicted cross section
by 20\% at low \ET~and 10\% at high \ET, while
the variation of the isolation requirement between 2 and 6~GeV changes
the predicted cross section by no more than 2\%.
The MSTW 2008 PDFs~\cite{mstw08} are used as a cross-check of
the choice of PDF.
The central values obtained with the MSTW 2008 PDFs are between 3 and
5\% higher than those predicted using the CTEQ 6.6 PDFs.

The NLO calculation provided by {\tt JETPHOX} predicts a cross section
at parton level, which does not include effects of hadronisation nor the
underlying event and pileup (\textit{i.e.} multiple proton-proton interactions
in the same bunch crossing).
The non-perturbative effects on the cross section due to hadronisation
are evaluated using the simulated {\tt PYTHIA} and {\tt HERWIG} 
signal samples described in Section~\ref{subsec:simulated_data},
by evaluating the ratio of the cross section before and after hadronisation 
and underlying event simulation.
The ratios are estimated to be within 1\% (2\%) of unity in {\tt PYTHIA} 
({\tt HERWIG}) for all \et\ and $\eta$ regions under study.
To account for the effect of the underlying event and pileup on the measured
isolation energy, a correction to the photon transverse isolation energy 
measured in data is applied, using a procedure described in Section~\ref{subsec:isolation}.

\section{Photon Reconstruction, Identification and Isolation}
\label{sec:PhotonID}
\subsection{Photon reconstruction and preselection}
\label{subsec:photon_reconstruction}
Photon reconstruction is seeded by clusters in the electromagnetic
calorimeter with transverse energies exceeding 2.5~GeV, measured in
projective towers of 3$\times$5 cells in $\eta\times\phi$ in the
second layer of the calorimeter.  An attempt is made to match these
clusters with tracks that are reconstructed in the inner detector and
extrapolated to the calorimeter.  Clusters without matching tracks are
directly classified as ``unconverted'' photon candidates.  Clusters
with matched tracks are considered as electron candidates.  To recover
photon conversions, clusters matched to pairs of tracks originating
from reconstructed conversion vertices in the inner detector are
considered as ``converted'' photon candidates.  To increase the
reconstruction efficiency of converted photons, conversion candidates
where only one of the two tracks is reconstructed (and does not have
any hit in the innermost layer of the pixel detector) are also
retained~\cite{Winter_confNote,ATLAS_CSC}.

The final energy measurement, for both converted and unconverted
photons, is made using only the calorimeter, with a cluster size that
depends on the photon classification.  In the barrel, a cluster
corresponding to 3$\times$5 ($\eta\times\phi$) cells in the second
layer is used for unconverted photons, while a cluster of 3$\times$7
($\eta\times\phi$) cells is used for converted photon candidates (to
compensate for the opening between the conversion products in the
$\phi$ direction due to the magnetic field).  In the end-cap, a
cluster size of 5$\times$5 is used for all candidates.  A dedicated
energy calibration~\cite{ATLAS_CSC} is then applied separately for
converted and unconverted photon candidates to account for upstream
energy loss and both lateral and longitudinal leakage.

Photon candidates with calibrated transverse energies ($\ET^\gamma$)
above 15 GeV are retained for the successive analysis steps.  To
minimise the systematic uncertainties related to the efficiency
measurement at this early stage of the experiment, the cluster
barycenter in the second layer of the electromagnetic calorimeter is
required to lie in the pseudorapidity region $|\eta^\gamma|< 1.37$, or
$1.52\leq |\eta^\gamma|<1.81$.  Photon candidates with clusters
containing cells overlapping with few problematic regions of the
calorimeter readout are removed.  After the above preselection, 1.3
million photon candidates remain in the data sample.

\subsection{Photon identification}
\label{subsec:photon_identification}
Shape variables computed from the lateral and longitudinal energy
profiles of the shower in the calorimeters are used to discriminate
signal from background.  The exact definitions of the discriminating
variables are provided in Appendix~\ref{app:IsEM-DV}.  Two sets of
selection criteria (denoted ``loose'' and ``tight'') are defined, each
based on independent requirements on several shape variables.  The
selection criteria do not depend on the photon candidate transverse
energy, but vary as a function of the photon reconstructed
pseudorapidity, to take into account variations in the total thickness
of the upstream material and in the calorimeter geometry.

\subsubsection{Loose identification criteria}
\label{subsubsec:photon_loose}
A set of loose identification criteria for photons is defined based on
independent requirements on three quantities:
\begin{itemize}
\item the leakage $\Rhad$ in the first layer of the hadronic
  compartment beyond the electromagnetic cluster, defined as the ratio
  between the transverse energy deposited in the first layer of the
  hadronic calorimeter and the transverse energy of the photon
  candidate;
\item the ratio \Reta~between the energy deposits in $3\times7$ and
  $7\times7$ cells in the second layer of the electromagnetic
  calorimeter;
\item the RMS width \wetatwo~of the energy distribution along $\eta$
  in the second layer of the electromagnetic calorimeter.
\end{itemize}
True prompt photons are expected to have small hadronic leakage
(typically below 1--2\%) and a narrower energy profile in the
electromagnetic calorimeter, more concentrated in the core of the
cluster, with respect to background photon candidates from jets.

The loose identification criteria on $\Rhad$, $\Reta$ and $\wetatwo$
are identical for converted and unconverted candidates.  They have
been chosen, using simulated prompt photon events, in order to obtain
a prompt photon efficiency, with respect to reconstruction, rising
from 97\% at $\ET^\gamma = 20~\GeV$ to above 99\% for $\ET^\gamma >
40~\GeV$ for both converted and unconverted
photons~\cite{Winter_confNote}.  The number of photon candidates in
data passing the preselection and loose photon identification criteria
is 0.8 million.

\subsubsection{Tight identification criteria}
\label{subsubsec:photon_tight}
To further reject the background, the selection requirements on the
quantities used in the loose identification are tightened. In
addition, the transverse shape along the $\phi$ direction in the
second layer (the variable $\Rphi$, computed from the ratio between
the energy deposits in $3\times3$ and $3\times7$ cells) and the shower
shapes in the first layer of the calorimeter are examined. Several
variables that discriminate single photon showers from overlapping
nearby showers (in particular those which originate from neutral meson
decays to photon pairs) are computed from the energy deposited in the
first layer:
\begin{itemize}
\item the total RMS width \wtot\ of the energy distribution along
  $\eta$;
\item the asymmetry \Eratio\ between the first and second maxima in
  the energy profile along $\eta$;
\item the energy difference \DeltaE\ between the second maximum and
  the minimum between the two maxima;
\item the fraction \Fside\ of the energy in seven strips centered (in
  $\eta$) around the first maximum that is not contained in the three
  core strips;
\item the RMS width \wthree\ of the energy distribution computed with
  the three core strips.
\end{itemize}

The first variable rejects candidates with wide showers consistent
with jets.  The second and third variables provide rejection against
cases where two showers give separated energy maxima in the first
layer.  The last two variables provide rejection against cases where
two showers are merged in a wider maximum.

The tight selection criteria are optimised independently for
unconverted and converted photons to account for the quite different
developments of the showers in each case.  They have been determined
using samples of simulated signal and background events prior to data
taking, aiming to obtain a average efficiency of 85\% with respect to
reconstruction for true prompt photons with transverse energies
greater than 20~GeV~\cite{Winter_confNote}.  About 0.2 million photon
candidates are retained in the data sample after applying the tight
identification requirements.

\subsection{Photon transverse isolation energy}
\label{subsec:isolation}
An experimental isolation requirement, based on the transverse energy
deposited in the calorimeters in a cone around the photon candidate,
is used in this measurement to identify isolated prompt photons and to
further suppress the main background from $\pi^0$ (or other neutral
hadrons decaying in two photons), where the $\pi^0$ is unlikely to
carry the full original jet energy.  The transverse isolation energy
(\Etiso) is computed using calorimeter cells from both the
electromagnetic and hadronic calorimeters, in a cone of radius 0.4 in
the $\eta-\phi$ space around the photon candidate. The contributions
from $5\times 7$ electromagnetic calorimeter cells in the $\eta-\phi$
space around the photon barycenter are not included in the sum.  The
mean value of the small leakage of the photon energy outside this
region, evaluated as a function of the photon transverse energy, is
subtracted from the measured value of \Etiso.  The typical size of
this correction is a few percent of the photon transverse energy.
After this correction, \Etiso\ for truly isolated photons is nominally
independent of the photon transverse energy.

In order to make the measurement of \Etiso\ directly comparable to
parton-level theoretical predictions, such as those described in
Section~\ref{sec:Theory}, \Etiso\ is further corrected by subtracting
the estimated contributions from the underlying event and from pileup.
This correction is computed on an event-by-event basis using a method
suggested in Refs.~\cite{Cacciari:area} and~\cite{Cacciari:UE}.  Based
on the standard seeds for jet reconstruction, which are
noise-suppressed three-dimensional topological
clusters~\cite{ATLAS_detector}, and for two different pseudorapidity
regions ($|\eta|<1.5$ and $1.5<|\eta|<3.0$), a $k_{T}$ jet-finding
algorithm~\cite{Ellis_kt,Catani_kt}, implemented in {\tt
  FastJet}~\cite{FastJet}, is used to reconstruct all jets without any
explicit transverse momentum threshold.  During reconstruction, each
jet is assigned an area via a Voronoi tessellation~\cite{Voronoi} of
the $\eta-\phi$ space. According to the algorithm, every point within
a jet's assigned area is closer to that jet than any other jet.  The
transverse energy density for each jet is then computed from the ratio
between the jet transverse energy and its area.  The ambient
transverse energy density for the event, from pileup and underlying
event, is taken to be the median jet transverse energy density.
Finally, this ambient transverse energy density is multiplied by the
area of the isolation cone to compute the correction to \Etiso.

The estimated ambient transverse energy fluctuates significantly
event-by-event, reflecting the fluctuations in the underlying event
and pileup activity in the data.  The mean correction to the
calorimeter transverse energy in a cone of radius $R=0.4$ for an event
with one $pp$ interaction is around 440 MeV in events simulated with
{\tt PYTHIA} and 550 MeV in {\tt HERWIG}.  In the data, the mean
correction is 540 MeV for events containing at least one photon
candidate with $\ET > 15$ GeV and exactly one reconstructed primary
vertex, and increases by an average of 170 MeV with each additional
reconstructed primary vertex.  The average number of reconstructed
primary vertices for the sample under study is 1.56.  The distribution
of measured ambient transverse energy densities for photons passing
the tight selection criteria is shown in
Fig.~\ref{fig:ambient_energy_density}.  The impact of this correction
on the measured cross section is discussed in
Section~\ref{subsec:syst_purity}.  For a consistent comparison of this
measurement to a theoretical prediction which incorporates an
underlying event model, the method described above should be applied
to the generated final state in order to evaluate and apply the
appropriate event-by-event corrections.

\begin{figure}
  \centering
  \includegraphics[width=1.0\columnwidth]{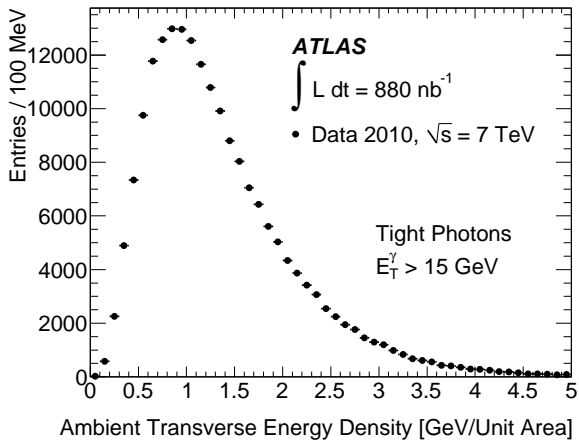}
  \caption{The measured ambient transverse energy densities, using the
    jet-area method, for events with at least one tight photon. The
    ambient transverse energy contains contributions from both the
    underlying event and pileup.  The broad distribution reflects the
    large event-to-event fluctuations.}
  \label{fig:ambient_energy_density}
\end{figure}

After the leakage and ambient-transverse-energy corrections, the
\Etiso\ distribution for direct photons in simulated events is
centered at zero, with an RMS width of around 1.5 GeV (which is
dominated by electronic noise in the calorimeter).  In the following,
all photon candidates with $\Etiso<3$ GeV are considered to be
experimentally isolated.  This criterion can be related to a cut on
the transverse isolation energy calculated at the parton level in {\tt
  PYTHIA}, in order to mimic the isolation criterion implemented in
{\tt JETPHOX}.  This parton-level isolation is the total transverse
energy of all partons that lie inside a cone of radius $R=0.4$ around
the photon direction and originated from the same quark emitting the
photon in either ISR or FSR.  The efficiency of the experimental
isolation cut at 3 GeV for photons radiated off partons in {\tt
  PYTHIA} is close to the efficiency of a parton-level isolation cut
at 4 GeV.  This cut on the parton-level isolation is equivalent to the
same cut on a particle-level isolation, which measures the transverse
energy in a cone of radius $R=0.4$ around the photon after
hadronisation (with the underlying event removed).  The experimental
isolation criterion is expected to reject roughly 50\% of background
candidates with transverse energy greater than 15~GeV.

About 110 thousand photon candidates satisfy the tight identification
criteria and have $\Etiso<3$ GeV: around 74 thousand are reconstructed
as unconverted photons and 36 thousand as converted photons.  The
transverse energy distribution of these candidates is shown in
Fig.~\ref{fig:et_distribution}. For comparison, the initial
distribution of all photon candidates after the reconstruction and
preselection is also shown.

\begin{figure}
  \centering
  \includegraphics[width=1.0\columnwidth]{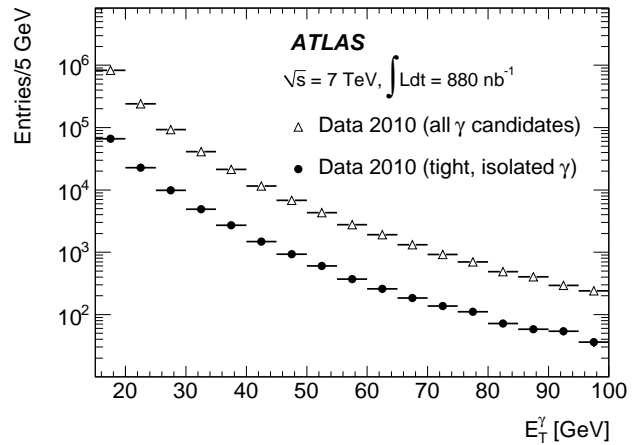}
  \caption{Transverse energy distribution of photon candidates
    selected in data, after reconstruction and preselection (open
    triangles) and after requiring tight identification criteria and
    transverse isolation energy lower than 3 GeV (full circles).  The
    photon candidates have pseudorapidity $|\eta^\gamma|<1.37$ or
    $1.52\leq|\eta^\gamma|<1.81$.}
  \label{fig:et_distribution}
\end{figure}

\section{Signal Efficiency}
\label{sec:Efficiency}
\subsection{Reconstruction and preselection efficiency}
\label{subsec:reco_eff}
The reconstruction and preselection efficiency, $\varepsilon_{\rm
  reco}$, is computed from simulated events as a function of the true
photon transverse energy for each pseudorapidity interval under study.
It is defined as the ratio between the number of true prompt photons
that are reconstructed -- after preselection -- in a certain
pseudorapidity interval and have reconstructed $\Etiso<3$ GeV, and the
number of true photons with true pseudorapidity in the same
pseudorapidity interval and with particle-level transverse isolation
energy lower than 4 GeV.  The efficiency of the requirement
$\ET^\gamma>15$ GeV for prompt photons of true transverse energy
greater than the same threshold is taken into account in
Section~\ref{sec:Unfolding}.

The reconstruction and preselection efficiencies are calculated using
a cross-section-weighted mixture of direct photons produced in
simulated $\gamma$-jet events and of fragmentation photons produced in
simulated dijet events.  The uncertainty on the reconstruction
efficiency due to the difference between the efficiency for direct and
fragmentation photons, and the unknown ratio of the two in the final
sample of selected signal photons, are taken into account as sources
of systematic uncertainty in Section~\ref{subsec:syst_photonreco}.

The average reconstruction and preselection efficiency for isolated
prompt photons with $|\eta^\gamma_{\rm true}|<1.37$ or
$1.52\leq|\eta^\gamma_{\rm true}|<1.81$ is around 82\%; the 18\%
inefficiency is due to the inefficiency of the reconstruction
algorithms at low photon transverse energy (a few \%), to the
inefficiency of the isolation requirement (5\%) and to the acceptance
loss from a few inoperative optical links of the calorimeter readout
\footnote{This inefficiency will be recovered in future data collected
  by ATLAS, as the faulty optical transmitters will be replaced during
  the LHC shutdown at the end of the 2010 run.}.

\subsection{Identification efficiency}
\label{subsec:offline_eff}
The photon identification efficiency, $\varepsilon_{\rm ID}$, is
similarly computed as a function of transverse energy in each
pseudorapidity region.  It is defined as the efficiency for
reconstructed (true) prompt photons, with measured $\Etiso<3$ GeV, to
pass the tight photon identification criteria described in
Section~\ref{subsec:photon_identification}.  The identification
efficiency is determined from simulation after shifting the photon
shower shapes by ``shower-shape correction factors'' that account for
the observed average differences between the discriminating variables'
distributions in data and MC.  The simulated sample used contains all
the main QCD signal and background processes. The average differences
between data and simulation are computed after applying the tight
identification criteria.  The typical size of the correction to the MC
efficiency is $-3\%$ and is always between $-5\%$ and zero.  The
photon identification efficiency after all selection criteria
(including isolation) are applied is shown in
Fig.~\ref{fig:offlineeff_photons} and in
Table~\ref{tab:offleff_all_photons}, including the systematic
uncertainties that are discussed in more detail in
Section~\ref{subsec:syst_photonreco}.  The efficiencies for converted
photons are, on average, 3-4\% lower than for unconverted photons with
the same pseudorapidity and transverse energy.
\begin{table}[!htb]
\centering
\caption{Isolated prompt photon identification efficiency
  in the intervals of the photon pseudorapidity and transverse energy 
  under study.
}
\label{tab:offleff_all_photons}
\footnotesize
\begin{tabular}{lccc}
\hline\hline
$\ET^\gamma$ & \multicolumn{3}{c}{Identification Efficiency} \\
$[{\rm GeV}]$& \multicolumn{3}{c}{$[\%]$} \\
\hline
             & $0.00 \leq |\eta^\gamma| < 0.60$ & $0.60 \leq |\eta^\gamma| < 1.37$ & $1.52 \leq |\eta^\gamma| < 1.81$ \\

\hline
   $[15,  20)$ & $63.3\pm 6.6$ & $63.5\pm 6.9$ & $72.2\pm 8.4$\\
   $[20,  25)$ & $73.5\pm 6.1$ & $73.5\pm 6.8$ & $81.6\pm 8.3$\\
   $[25,  30)$ & $80.2\pm 5.4$ & $80.8\pm 5.7$ & $86.7\pm 6.6$\\
   $[30,  35)$ & $85.5\pm 4.5$ & $85.3\pm 4.8$ & $90.4\pm 5.9$\\
   $[35,  40)$ & $85.2\pm 3.9$ & $89.3\pm 4.3$ & $92.3\pm 5.0$\\
   $[40,  50)$ & $89.2\pm 3.3$ & $92.1\pm 3.6$ & $93.5\pm 4.6$\\
   $[50,  60)$ & $91.3\pm 3.1$ & $94.1\pm 2.8$ & $93.9\pm 3.6$\\
   $[60, 100)$ & $92.2\pm 2.6$ & $94.8\pm 2.6$ & $94.2\pm 2.9$\\
\hline\hline
\end{tabular}
\end{table}

\begin{figure}
  \centering
  \includegraphics[width=1.0\columnwidth]{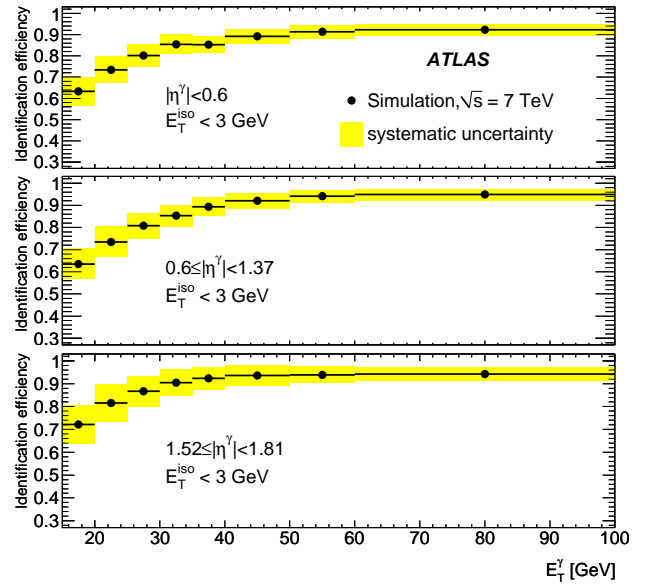}
  \caption{Efficiency of the tight identification criteria as a
    function of the reconstructed photon transverse energy for prompt
    isolated photons.  Systematic uncertainties are included.}
  \label{fig:offlineeff_photons}
\end{figure}

As a cross-check, photon identification efficiencies are also inferred
from the efficiencies of the same identification criteria applied to
electrons selected in data from $W$ decays.  Events containing $W\to
e\nu$ candidates are selected by requiring: a missing transverse
energy greater than 25 GeV (corresponding to the undetected neutrino);
an opening azimuthal angle larger than 2.5 radians between the missing
transverse energy vector and any energetic jets ($\ET>15$ GeV) in the
event; an electron transverse isolation energy in a cone of radius 0.4
in the $\eta-\phi$ space smaller than 0.3 times the electron
transverse momentum; and a track, associated to the electron, that
passes track-quality cuts, such as a large amount of transition
radiation produced in the TRT and the presence of hits in the silicon
trackers.  These selection criteria, which do not rely on the shape of
the electron shower in the calorimeter, are sufficient to select a
$W\to e\nu$ sample with a purity greater than 95\%.  The
identification efficiency of converted photons is taken from the
efficiency for selected electrons to pass the tight photon selection
criteria. This approximation is expected to hold to within 3\% from
studies of simulated samples of converted isolated prompt photons and
of isolated electrons from $W$ decays.  For unconverted photons, the
electrons in data are used to infer shower-shape corrections.  These
corrections are then applied to unconverted photons in simulation, in
order to calculate the unconverted photon efficiency from Monte Carlo.
The results from the electron extrapolation method are consistent with
those from the simulation, with worse precision due to the limited
statistics of the selected electron sample.

\subsection{Trigger efficiency}
\label{subsec:trig_eff}
The efficiency of the calorimeter trigger, relative to the photon
reconstruction and identification selection, is defined as the
probability for a true prompt photon, passing the tight photon
identification criteria and with $\Etiso<3$ GeV, to pass the trigger
selection.  It is estimated in two steps. First, using a prescaled
sample of minimum bias triggers, the efficiency of a lower threshold
($\approx 3.5$ GeV) level-1 calorimeter trigger is determined.  The
measured efficiency of this trigger is 100\% for all photon candidates
with reconstructed $\ET^{\gamma}>15$ GeV passing tight identification
criteria.  Then, the efficiency of the high-level trigger is measured
using the sample of events that pass the level-1 calorimeter trigger
with the 3.5 GeV threshold.

The trigger efficiency for reconstructed photon candidates passing
tight selection criteria, isolated and with $\ET^{\gamma}>15$~\GeV~is
found to be $\varepsilon^{\rm trig}=(99.5\pm 0.5)\%$, constant within
uncertainties over the full $\ET$ and $\eta$ ranges under study.  The
quoted uncertainty is obtained from the estimation of the possible
bias introduced by using photon candidates from data, which are a
mixture of signal and background photon candidates.  Using Monte Carlo
samples the absolute difference of the trigger efficiency for a pure
signal sample and for a pure background sample is found to be smaller
than 0.5\% for isolated tight photon candidates with $\ET^{\gamma}>15$
GeV.

A comparison between the high-level trigger efficiency in data and in
the background predicted by the simulation is
shown in Fig.~\ref{fig:trigger_efficiency_datamc}. 

\begin{figure}
  \begin{center}
    \includegraphics[width=1.0\columnwidth]{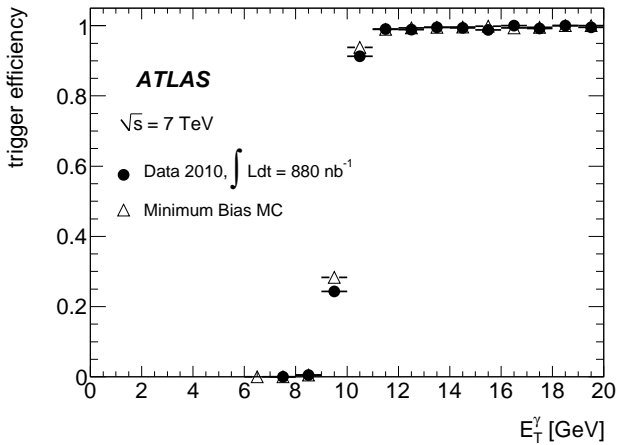}  
    \caption{Photon trigger efficiency, with respect to reconstructed
      isolated photon passing the tight identification criteria, as
      measured in data (circles) and simulated background events
      (triangles).}
    \label{fig:trigger_efficiency_datamc}
  \end{center}
\end{figure}

\section{Background subtraction and signal yield determination}
\label{sec:Purity}
A non-negligible residual contribution of background candidates is
expected in the selected photon sample, even after the application of
the tight identification and isolation requirements. Two methods are
used to estimate the background contribution from data and to measure
the prompt photon signal yield. The first one is used for the final
cross section measurement, while the second one is used as a cross
check of the former.  All estimates are made separately for
each region of pseudorapidity and transverse energy.

\subsection{Isolation vs. identification sideband counting method}
\label{sec:purity_2dsideband}
The first technique for measuring the prompt photon yield uses the
number of photon candidates observed in the sidebands of a
two-dimensional distribution to estimate the amount of background in
the signal region.  The two dimensions are defined by the transverse
isolation energy $\ET^{\rm iso}$ on one axis, and the photon
identification ($\gamma_{\rm ID}$) of the photon candidate on the
other axis.  On the isolation axis, the signal region contains photon
candidates with $\Etiso<3$ GeV, while the sideband contains photon
candidates with $\Etiso>5$ GeV.  On the other axis, photon candidates
passing the tight identification criteria (``tight'' candidates)
belong to the $\gamma_{\rm ID}$ signal region, while those that fail
the tight identification criteria but pass a background-enriching
selection (``non-tight'' candidates) belong to the $\gamma_{\rm ID}$
sideband.  The non-tight selection requires photon candidates to fail
at least one of a subset of the photon tight identification criteria,
but to pass all criteria not in that subset.  All the shower shape
variables based on the energy measurement in the first layer of the
electromagnetic calorimeter are used to define the background
enriching selection, with the exception of $\wtot$, since it is found
to be significantly correlated with the \Etiso\ of background photon
candidates, while the photon yield measurement relies on the
assumption of negligible (or small) correlations between the
transverse isolation energy and the shower shape quantities used to
define the background enriching selection.

The signal region (region ``A'') is therefore defined by photon
candidates passing the tight photon identification criteria and having
experimental $\Etiso<3$ GeV.  The three background control regions
consist of photon candidates either:
\begin{itemize}
\item passing the tight photon identification criteria but having
  experimental $\Etiso>5$ GeV (region ``B'')
\item having $\Etiso<3$ GeV and passing the background-enriching
  identification criteria (region ``C'')
\item having $\Etiso>5$ GeV and passing the background-enriching
  identification criteria (region ``D'').
\end{itemize}
A sketch of the two dimensional plane and of the signal and background
control region definitions is shown in Fig.~\ref{fig:2dplane}.

\begin{figure}[!htb]
  \centering    
  \includegraphics[width=0.9\columnwidth]{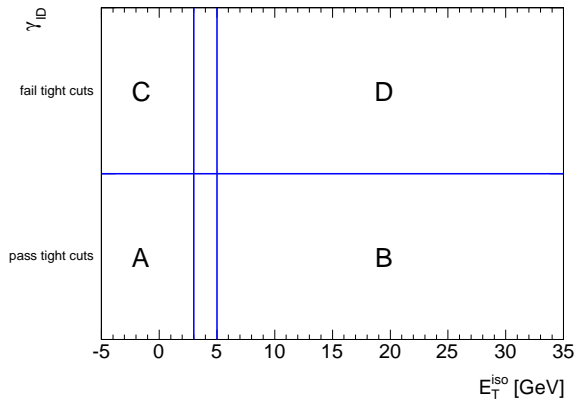}
  \caption{Illustration of the two-dimensional plane, defined by means
    of the transverse isolation energy and a subset of the photon
    identification (ID) variables, used for estimating, from the
    observed yields in the three control regions ($B, C, D$), the
    background yield in the signal region ($A$).}
  \label{fig:2dplane}
\end{figure}

The method assumes that the signal contamination in the three
background control regions is small, and that the isolation profile in
the non-tight regions is the same as that of the background in the
tight regions.  If these assumptions hold, then the number of
background candidates in the signal region can be calculated by taking
the ratio of candidates in the two non-tight regions ($N_C/N_D$), and
multiplying it by the number of candidates in the tight, non-isolated
region ($N_B$).  The number of isolated prompt photons passing the
tight identification criteria is therefore:
\begin{equation}
  \label{eq:sigyield_nocorr}
  N_A^{\rm sig} = N_A - N_B\frac{N_C}{N_D}\ ,
\end{equation}
where
$N_A$ is the observed number of photon candidates in the signal region.

The assumption that the signal contamination in the background control
regions is small is checked using prompt photon MC samples.  As the
number of signal events in the background control regions is always
positive and non-zero, corrections are applied to limit the effects on
the final result.  For this purpose, Eq.~\ref{eq:sigyield_nocorr} is
modified in the following way:
\begin{equation}
  \label{eq:sigyield_leakage}
  N_A^{\rm sig} = N_A - (N_B-c_BN_A^{\rm sig})\frac{(N_C-c_CN_A^{\rm sig})}{(N_D-c_DN_A^{\rm sig})}\ ,
\end{equation}
where $c_K\equiv\frac{N^{\rm sig}_K}{N^{\rm sig}_A}$ (for $K \in
\{B,C,D\}$) are the signal leakage fractions extracted from
simulation.  Typical values for $c_B$ are between 3\% and 17\%,
increasing with the photon candidate transverse energy; for $c_C$,
between 2\% and 14\%, decreasing with $\ET^\gamma$.  $c_D$ is always
less than 2\%.  The total effect of these corrections on the measured
signal photon purities is typically less than 5\%.

The isolated prompt photon fraction measured with this method, as a
function of the photon reconstructed transverse energy, is shown in
Fig.~\ref{fig:purity_sideband}. The numbers of isolated prompt photon
candidates measured in each pseudorapidity and transverse energy
interval are also reported in Table~\ref{tab:yields}.  The systematic
uncertainties on the measured prompt photon yield and fraction in the
selected sample are described in Section~\ref{subsec:syst_purity}.

\begin{figure}
  \centering    
  \includegraphics[width=1.0\columnwidth]{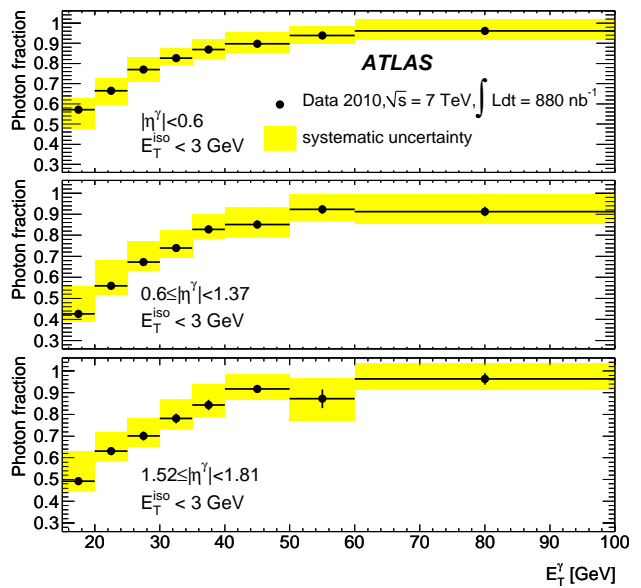}
  \caption{Fraction of isolated prompt photons as a function of the
    photon transverse energy, as obtained with the two-dimensional
    sideband method.}
  \label{fig:purity_sideband}
\end{figure}

\begin{table*}[!htb]
\centering
\caption{Observed number of isolated prompt photons in the photon transverse 
  energy and pseudorapidity intervals under study. The first uncertainty is 
  statistical, the second is the systematic uncertainty, evaluated as described 
  in Section~\ref{subsec:syst_purity}.}
\label{tab:yields}
\footnotesize
\renewcommand{\arraystretch}{1.5}
\begin{tabular}{lrrrrrrrrr}
\hline\hline
& \multicolumn{9}{c}{Isolated prompt photon yield} \\
\hline
$\ET^\gamma~[{\rm GeV}]$                & \multicolumn{3}{c}{$0.00 \leq |\eta^\gamma| < 0.60$} & \multicolumn{3}{c}{$0.60 \leq |\eta^\gamma| < 1.37$} &  \multicolumn{3}{c}{$1.52 \leq |\eta^\gamma| < 1.81$} \\
\hline
  $[15,  20)$ &  (119 & $\pm     3$ & $^{+  12}_{- 20})\times 10^2$  & (130 & $\pm    4$ & $^{+  40}_{-  11})\times 10^2$  &  (72 & $\pm    2$ & $^{+  20}_{-   7})\times 10^2$   \\
  $[20,  25)$ &  (501 & $\pm    12$ & $^{+47}_{-53})\times 10^1   $  & (578 & $\pm   18$ & $^{+ 125}_{-  45})\times 10^1$  &  (304 & $\pm    10$ & $^{+  40}_{-  23})\times 10^1$   \\
  $[25,  30)$ &  (260 & $\pm     7$ & $^{+20}_{-21})\times 10^1   $  & (306 & $\pm   10$ & $^{+  46}_{-  18})\times 10^1$  &  (135 & $\pm     6$ & $^{+  16}_{-  10})\times 10^1$   \\
  $[30,  35)$ &  (146 & $\pm     5$ & $^{+9}_{-6})\times 10^1     $  & (160 & $\pm    6$ & $^{+  19}_{-   9})\times 10^1$  &  (73 & $\pm     4$ & $^{+    8}_{-    5})\times 10^1$   \\
  $[35,  40)$ &  (82  & $\pm     4$ & $^{+5}_{-4})\times 10^1     $  & (102 & $\pm    4$ & $^{+   9}_{-   6})\times 10^1$  &  (44 & $\pm     3$ & $^{+    5}_{-    3})\times 10^1$   \\
  $[40,  50)$ &  (77  & $\pm     3$ & $^{+5}_{-4})\times 10^1     $  & (98 & $\pm     4$ & $^{+   9}_{-   7})\times 10^1$  &  (38 & $\pm     2$ & $^{+    3}_{-    2})\times 10^1$   \\
  $[50,  60)$ &  (329 & $\pm    20$ & $^{+17}_{-14})\times 10^0   $  & (420 & $\pm   20$ & $\pm 30        )\times 10^0$  &  (147 & $\pm     16$ & $^{+    16}_{-    17})\times 10^0$   \\
  $[60, 100)$ &  (329 & $\pm    20$ & $^{+19}_{-15})\times 10^0   $  & (370 & $\pm   20$ & $^{+  30}_{-  20})\times 10^0$  &  (154 & $\pm     12$ & $^{+    12}_{-     8})\times 10^0$   \\
\hline\hline
\end{tabular}
\end{table*}

\subsection{Isolation template fit method}
The second method relies on a binned maximum likelihood fit to the
\Etiso\ distribution of photon candidates selected in data which pass
the tight identification criteria. The distribution is fit to the sum
of a signal template and a background template, determined from
control samples extracted from data.  This is similar to the technique
employed in~\cite{CDF_promptphoton}, but relies less on simulation for
signal and background templates.  The signal template is determined
from the \Etiso\ distribution of electrons from $W$ and $Z$ decays,
selected using the criteria described in~\cite{WZpaper}.  Electrons
from $W$ decays are required to fulfill tight selection criteria on
the shapes of their showers in the electromagnetic calorimeter and to
pass track-quality requirements, including the presence of
transition-radiation hits. They must also be accompanied by $\met>25$
GeV, and the electron--$\met$ system must have a transverse mass
larger than $40$ GeV.  Electrons from $Z$ decays are selected with
looser criteria, but the pair must have an invariant mass close to the
$Z$ mass. A single signal template is constructed for each region in
$|\eta|$, exploiting the independence of \Etiso\ from the transverse
energy of the object (after applying the corrections described in
Section~\ref{subsec:isolation}) to maximize the available statistics.
A small bias is expected due to differences between the electron and
photon \Etiso\ distributions, especially in regions where there is
significant material upstream of the calorimeter.  A shift of the
signal template is applied to the electron distributions extracted
from data to compensate for the differences between electrons and
photons seen in simulation. This shift, computed using simulated
photon and electron samples, increases from 100 MeV to 600 MeV with
increasing $|\eta^\gamma|$.  The background template is extracted from
data for each (\ET, $|\eta|$) bin, using the same reverse-cuts
procedure as in the two-dimensional sideband technique. A
simulation-based correction, typically of the order of 3-4\%, is
applied to the final photon fraction to account for signal which leaks
into the background template.  The fit is performed in each region of
$|\eta^\gamma|$ for the individual bins in transverse energy, and the
signal yield and fraction are extracted. An example of such a fit is
shown in Fig.~\ref{fig:2tf_data_example}. The results from this
alternative technique are in good agreement with those from the
simpler counting method described in the previous subsection, with
differences typically smaller than 2\% and within the systematic
uncertainties that are uncorrelated between the two methods.

\begin{figure}
  \centering
  \includegraphics[width=1.0\columnwidth]{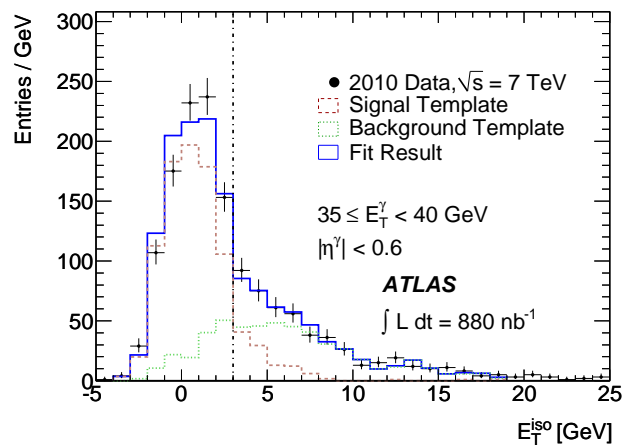}
  \caption{Example of a fit to extract the fraction of prompt photons
    using the isolation template technique in the region $0 \leq
    |\eta| < 0.6$ and $35 \leq \ET^{\gamma} < 40\GeV$.  The signal
    template is derived from electrons selected from $W$ or $Z$
    decays, and is shown with a dashed line.  The background template
    is derived from a background-enriched sample, and is represented
    by a dotted line. The estimated photon fraction is 0.85 and its
    statistical uncertainty is 0.01.  }
  \label{fig:2tf_data_example}
\end{figure}

\subsection{Electron background subtraction}
\label{sec:electron_bkg}


The background of prompt electrons misidentified as photons needs also
to be considered. The dominant electron production mechanisms are
semileptonic hadron decays (mostly from hadrons containing heavy
flavor quarks) and decays of electroweak bosons (the largest
contribution being from $W$ decays).  Electrons from the former are
often produced in association with jets, and have \Etiso\ profiles
similar to the dominant backgrounds from light mesons. They are
therefore taken into account and subtracted using the two-dimensional
sideband technique described in Section~\ref{sec:purity_2dsideband}.
Conversely, electrons from $W$ and $Z$ decays have \Etiso\ profiles
that are similar to those of signal photons.  The contribution of this
background to the signal yield computed in
Section~\ref{sec:purity_2dsideband} needs therefore to be removed
before the final measurement of the cross section.

The fraction of electrons reconstructed as photon candidates is
estimated from the data, as a function of the electron transverse
energy and pseudorapidity, using a control sample of $Z\to e^+e^-$
decays.  The average electron misidentification probability is around
8\%. Using the $W\rightarrow~e\nu$ and $Z\rightarrow~ee$ cross section
times branching ratio measured by ATLAS in $pp$ collisions at
$\sqrt{s}=7$ TeV~\cite{WZpaper}, the estimated fraction of photon
candidates due to isolated electrons is found to be on average
$\sim0.5\%$, varying significantly with transverse energy.  A maximum
contamination of $(2.5\% \pm 0.8\%)$ is estimated for transverse
energies between 40 and 50 GeV, due to the kinematic distribution of
electrons from $W$ and $Z$ decays. The uncertainties on these
estimates are less than 1\% of the photon yield.

\section{Cross section measurement}
\label{sec:Unfolding}
The differential cross section is measured by computing:

\begin{equation}
  \frac{d\sigma}{dE_{\rm T}^{\gamma}} = \frac{N_{\mathrm{yield}}~U}{\left(\int{\mathcal{L}dt}\right)~\Delta{E_{\rm T}^{\gamma}}~\varepsilon_{\mathrm{trigger}}~\varepsilon_{\mathrm{reco}}~\varepsilon_{\mathrm{ID}}}\ .
\end{equation}

The observed signal yield ($N_{\mathrm{yield}}$) is divided by the
widths of the \ET-intervals ($\Delta{E_{\rm T}^{\gamma}}$) and by the
product of the photon identification efficiency
($\varepsilon_{\mathrm{ID}}$, determined in
Section~\ref{subsec:offline_eff}) and of the trigger efficiency
relative to photon candidates passing the identification criteria
($\varepsilon_{\mathrm{trigger}}$, determined in
Section~\ref{subsec:trig_eff}).  The spectrum obtained this way, which
depends on the reconstructed transverse energy of the photon
candidates, is then corrected for detector energy resolution and
energy scale effects using bin-by-bin correction factors (the
``unfolding coefficients'' $U$) evaluated using simulated samples.
The corrected spectrum, which is then a function of the true photon
energy, is divided by the photon reconstruction efficiency
$\varepsilon_{\mathrm{reco}}$ (Section~\ref{subsec:reco_eff}) and by
the integrated luminosity of the data sample, $\int{\mathcal{L}dt}$.

The unfolding coefficients are evaluated from the ratio of the true to
reconstructed \ET~distributions of photon candidates, using {\tt
  PYTHIA} isolated prompt photon simulated samples.  This procedure is
justified by the small bin-to-bin migrations (typically of the order
of a few \%) that are expected, given the good electromagnetic
calorimeter energy resolution compared to the width of the transverse
energy intervals used in this analysis (between 5 and 40 GeV).  The
values of the unfolding coefficients are slightly higher than 1 and
decrease as a function of \ET, approaching 1.  They differ from 1 by
less then 2\% in the $|\eta^\gamma|$ region between 0.0 and 0.6, and
by less than 5-7\% in the other two $|\eta^\gamma|$ regions, where
more material upstream of the electromagnetic calorimeter is present.

\section{Systematic Uncertainties}
\label{sec:Systematics}
Several sources of systematic uncertainties on the cross section are
identified and evaluated as described in the following sections.  The
total systematic uncertainty is obtained by combining the various
contributions, taking into account their correlations: uncorrelated
uncertainties are summed in quadrature while a linear sum of
correlated uncertainties is performed.

\subsection{Reconstruction, identification, trigger efficiencies}
\label{subsec:syst_photonreco}
The systematic uncertainty on the reconstruction efficiency from the
experimental isolation requirement is evaluated from the prompt photon
simulation varying the value of the isolation criterion by the average
difference (of the order of 500 \MeV) observed for electrons between
simulation and data control samples. It is 2.5\% in the pseudorapidity
regions covered by the barrel calorimeter and 4.5\% in the end-caps.

The systematic uncertainty on the identification efficiency
due to the photon shower-shape corrections is divided into two parts.
The first term evaluates the impact of treating the differences between
the distributions of the shower shape variables in data and simulation
as an average shift. This uncertainty is evaluated in the following way:
\begin{itemize}
\item A modified description of the detector material is used to
  produce a second sample of simulated photon candidates.  These
  candidates have different shower-shape distributions, due to the
  different amount of material upstream of and within the calorimeter.
  This alternative model contains an additional 10\% of material in
  the inactive volumes of the inner detector and 10\% of radiation
  length in front of the electromagnetic calorimeter.  This model is
  estimated to represent a conservative upper limit of the additional
  detector material that is not accounted for by the nominal
  simulation.
\item The correction procedure is applied to the nominal simulation to
  estimate the differences between the nominal and the alternative
  simulation. The shifts between the discriminating variable
  distributions in the nominal and the alternative simulation are
  evaluated, and are used to correct the shower shape variable
  distributions of the nominal simulation.
\item The photon efficiency from the nominal simulation is recomputed
  after applying these corrections, and compared with the efficiency
  obtained from the alternative simulation.
\end{itemize}
The difference between the efficiency estimated from the nominal
simulation (after applying the corrections) and the efficiency
measured directly in the alternative sample (with no corrections)
ranges from 3\% at $\ET^\gamma\sim 20\GeV$ to less than 1\% at
$\ET^\gamma\sim 80\GeV$.

The second part of the systematic uncertainty on the identification
efficiency accounts for the uncertainty on the extracted shower-shape
correction factors.  The correction factors were extracted by
comparing tight photons in data and simulation; to evaluate the
uncertainty associated with this choice, the same correction factors
are extracted using loose photons.  The difference in the final
efficiency when applying the tight corrections and the loose
corrections is then taken as the uncertainty.  This uncertainty drops
from 4\% to 1\% with increasing $\ET^\gamma$.

Additional systematic uncertainties that may affect both the
reconstruction and the identification efficiencies are evaluated
simultaneously for the product of the two, to take into account
possible correlations. These sources of uncertainty include the amount
of material upstream of the calorimeter; the impact of pile-up; the
relative fraction of direct and fragmentation photons in data with
respect to simulation; the misidentification of a converted photon as
unconverted; the difference between the {\tt PYTHIA} and {\tt HERWIG}
simulation models; the impact of a sporadic faulty calibration of the
cell energies in the electromagnetic calorimeter; and the imperfect
simulation of acceptance losses due to inoperative readout links in
the calorimeter.

Of all the uncertainties which contribute to this measurement, the
largest ones come from the uncertainty on the amount of material
upstream of the calorimeter (absolute uncertainties ranging between
1\% and 8\% and are larger at low $\ET^\gamma$), and from the
uncertainty on the identification efficiency due to the photon
shower-shape corrections (the absolute uncertainties are in the range
1-5\%, and are larger at low $\ET^\gamma$).

The uncertainty on the trigger efficiency, evaluated as described in
Section~\ref{subsec:trig_eff}, is 0.5\% and is nearly negligible
compared to all other sources.

\subsection{Signal yield estimates}
\label{subsec:syst_purity}

The following sources of systematic uncertainties affecting the
accuracy of the signal yield measurement using the two-dimensional
sideband technique are considered.

\subsubsection{Background isolation control region definition}

The signal yield is evaluated after changing the isolation control
region definition. The minimum value of \Etiso\ required for
candidates in the non-isolated control regions, which is set to 5 GeV
in the nominal measurement, is changed to 4 and 6 GeV. This check is
sensitive to uncertainties in the contribution of prompt photons from
QED radiation from quarks: these photons are less isolated than those
originating from the hard process.  Alternative measurements are also
performed where a maximum value of \Etiso\ is set to 10 or 15 GeV for
candidates in the non-isolated control regions, in order to reduce the
correlation between the isolation variable and the shower shape
distributions seen in simulated events for candidates belonging to the
upper tail of the isolation distribution. The largest positive and
negative variations of the signal yield with respect to the nominal
result are taken as systematic uncertainties. The signal photon
fraction changes by at most $\pm 2\%$ in all the transverse energy and
pseudorapidity intervals.

\subsubsection{Background photon identification control region definition}

The measurement is repeated reversing the tight identification
criteria on a number of strip variables ranging between two (only
$F_{\rm side}$ and $w_{s3}$) and five (all the variables based on the
first layer of the electromagnetic calorimeter).  The largest positive
and negative variations of the signal yield (with respect to the
nominal result) from these three alternative measurements are taken as
systematic uncertainties.  The effect on the signal photon fraction
decreases with increasing photon transverse energy, and is around 10\%
for $\ET^\gamma$ between 15 and 20 GeV.

\subsubsection{Signal leakage into the photon identification background control region}

From the photon identification efficiency studies, an upper limit of
5\% is set on the uncertainty on the fraction $c_C$ of signal photons
passing all the tight photon identification criteria except those used
to define the photon identification control region.  The signal yields
in each $\ET^\gamma, |\eta^\gamma|$ interval are thus measured again
after varying the estimated signal contamination in the photon
identification control regions ($c_C$ and $c_D/c_B$) by this
uncertainty, and the difference with the nominal result is taken as a
systematic uncertainty.  The signal fraction variations are always
below 6\%.

\subsubsection{Signal leakage into the isolation background control region}

The fractions $c_B$ and $c_D$ of signal photons contaminating the
isolation control regions depend on the relative amount of direct and
fragmentation photons in the signal selected in a certain $\ET^\gamma,
|\eta^\gamma|$ interval, since the latter are characterized by larger
nearby activity, and therefore usually have slightly larger transverse
isolation energies.  In the nominal measurement, the values of $c_B$
and $c_D$ are computed with the relative fractions of direct and
fragmentation photons predicted by {\tt PYTHIA}.  A systematic
uncertainty is assigned by repeating the measurement after varying
these fractions between 0\% and 100\%. The measured signal photon
fraction varies by less than 5\%.

\subsubsection{Signal photon simulation}

The signal yield is estimated using samples of prompt photons
simulated with {\tt HERWIG} instead of {\tt PYTHIA} to determine the
fraction of signal leaking into the three background control
regions. The variations of the signal photon fractions in each
$\ET^\gamma, |\eta^\gamma|$ interval are below 2\%.

\subsubsection{Correlations between the isolation and the photon
  identification variables for background candidates}

Non-negligible correlations between the isolation variable and the
photon identification quantities would affect
Eq.~\ref{eq:sigyield_leakage}: the true number of isolated tight
prompt photon candidates would be
\begin{equation}
  \label{eq:sigyield_correlation}
  N_A^{\rm sig}{=}N_A{-}R^{\rm bkg}(N_B{-}c_BN_A^{\rm sig})\frac{(N_C{-}c_CN_A^{\rm sig})}{(N_D{-}c_DN_A^{\rm sig})}
\end{equation}
where $R^{\rm bkg} \equiv \frac{N^{\rm bkg}_{A} N^{\rm
    bkg}_{D}}{N^{\rm bkg}_{B} N^{\rm bkg}_{C} }$ would then be
different from unity.  The simulation of background events shows a
small but non-negligible correlation between the isolation and the
discriminating shower shape variables used to define the photon
identification signal and background control regions.  The signal
yields are therefore recomputed with the formula in
Eq.~\ref{eq:sigyield_correlation}, using for $R^{\rm bkg}$ the value
predicted by the {\tt PYTHIA} background simulation, and compared with
the nominal results.  The effect is smaller than $0.6\%$ in the
$|\eta^\gamma|<1.37$ intervals and around $3.6\%$ for
$1.52\leq|\eta^\gamma|<1.81$.

\subsubsection{Transverse isolation energy corrections}

The effects of the \Etiso\ correction for the underlying event on the
estimated signal yield are also investigated.  As this is an
event-by-event correction, it cannot be unfolded from the observed
cross section.  The impact of this correction is evaluated by
estimating the signal yield, with and without the correction applied,
for events with only one reconstructed primary vertex (to eliminate
any effects of pileup).  The estimated signal yields using the
uncorrected values of \Etiso, normalized to the yields derived using
the corrected values, show no trend in $\ET^\gamma$ or $\eta$.
Furthermore, the impact on the cross section of the event-by-event
corrections is equivalent to that of an average correction of 540 MeV
applied to the transverse isolation energies of all photon candidates.
Similar studies in {\tt PYTHIA} and {\tt HERWIG} MC yield identical
results.

\subsection{Unfolding coefficients}
\label{subsec:syst_unfolding}

The unfolding coefficients used to correct the measured cross section
for \ET~bin-by-bin migrations are computed using simulated samples.
There are three sources of uncertainties on these coefficients.

\subsubsection{Energy scale uncertainty}

The uncertainty on the energy scale was estimated to be $\pm3\%$ in
test beam studies~\cite{testbeam}, and is confirmed to be below this
value from the comparison of the $Z\to e^+e^-$ invariant mass peak in
data and Monte Carlo.  The unfolding coefficients are thus recomputed
using simulated signal events where the true photon energy is shifted
by $\pm 3\%$.  The coefficients change by $\pm 10\%$. This uncertainty
introduces a relative uncertainty of about 10\% on the measured cross
section which is fully correlated between the different $\ET^\gamma$
intervals within each pseudorapidity range.

\subsubsection{Energy resolution uncertainty}

The uncertainty on the energy resolution may affect bin-by-bin
migrations between adjacent \ET~bins.  Test beam studies indicate
agreement between the sampling term of the resolution between data
simulation within 20\% relative. Furthermore, studies of the $Z\to
e^+e^-$ invariant mass distribution in data indicate that the constant
term of the calorimeter energy resolution is below 1.5\% in the barrel
and 3.0\% in the end-cap (it is 0.7\% in the simulation).  The
unfolding coefficients are thus recomputed after the reconstructed
energy of simulated photons smearing to take into account a 20\%
relative increase of the sampling term and a constant term of 1.5\% in
the barrel and 3.0\% in the end-cap.  The resulting variation of the
unfolding coefficients is always less than 1\%.  The uncertainty
arising from non-gaussian tails of the energy resolution function is
estimated by recomputing the coefficients using a prompt photon
simulation where a significant amount of material is added to the
detector model.  The variations of the unfolding coefficients are
smaller than 1\% in all the pseudorapidity and transverse energy
intervals under study.

\subsubsection{Simulated photon transverse energy distribution}

The unfolding coefficients, computed in $\ET^\gamma$~intervals of
non-negligible size, depend on the initial $\ET^\gamma$~distribution
predicted by {\tt PYTHIA}.  An alternative unfolding
technique~\cite{bayes_unfolding} is therefore used, which relies on
the repeated application of Bayes' theorem to iteratively obtain an
improved estimate of the unfolded spectrum.  This technique relies
less on the simulated original \ET~distribution of the prompt photons.
The differences between the cross-sections estimated using the
bin-by-bin unfolding and the iterative Bayesian unfolding are within
2\%, and are taken into account as an additional systematic
uncertainty.

\subsection{Luminosity}
\label{subsec:syst_luminosity}

The integrated luminosity is determined for each run by measuring
interaction rates using several ATLAS subdetectors at small angles to
the beam line, with the absolute calibration obtained from beam
position scans~\cite{ATLAS-CONF-2010-060}.  The relative systematic
uncertainty on the luminosity measurement is estimated to be 11\% and
translates directly into a 11\% relative uncertainty on the cross
section.

\section{Results and Discussion}
\label{sec:CrossSection}
The measured inclusive isolated prompt photon production cross
sections $d\sigma/d\ET^\gamma$ are shown in
Fig.~\ref{fig:xsec_measured_vs_jetphox_2dsideband_etabin000-060},
\ref{fig:xsec_measured_vs_jetphox_2dsideband_etabin060-137}, and
\ref{fig:xsec_measured_vs_jetphox_2dsideband_etabin152-181}.  They are
presented as a function of the photon transverse energy, for each of
the three considered pseudorapidity intervals.  They are also
presented in tabular form in Appendix~\ref{app:CS-tables}.  The
measurements extend from $\ET^\gamma=15$ GeV to $\ET^\gamma=100$ GeV
spanning almost three orders of magnitude.  The data are compared to
NLO pQCD calculations, obtained with the {\tt JETPHOX} program as
described in Section~\ref{sec:Theory}.  The error bars on the data
points represent the combination of the statistical and systematic
uncertainties (summed in quadrature): systematic uncertainties
dominate over the whole considered kinematic range.  The contribution
from the luminosity uncertainty (11\%) is shown separately (dotted
bands) as it represents a possible global offset of all the
measurements.  The total systematic uncertainties on the theoretical
predictions are represented with a solid band.  They are obtained by
summing in quadrature the contributions from the scale uncertainty,
the PDF uncertainty (at 68\% C.L.) and the uncertainty associated with
the choice of the parton-level isolation criterion.  The same
quantities are also shown, in the bottom panels of
Fig.~\ref{fig:xsec_measured_vs_jetphox_2dsideband_etabin000-060},
\ref{fig:xsec_measured_vs_jetphox_2dsideband_etabin060-137}, and
\ref{fig:xsec_measured_vs_jetphox_2dsideband_etabin152-181}, after
having been normalized to the expected NLO pQCD cross sections.

\begin{figure}
  \centering
  \includegraphics[width=1.0\columnwidth]{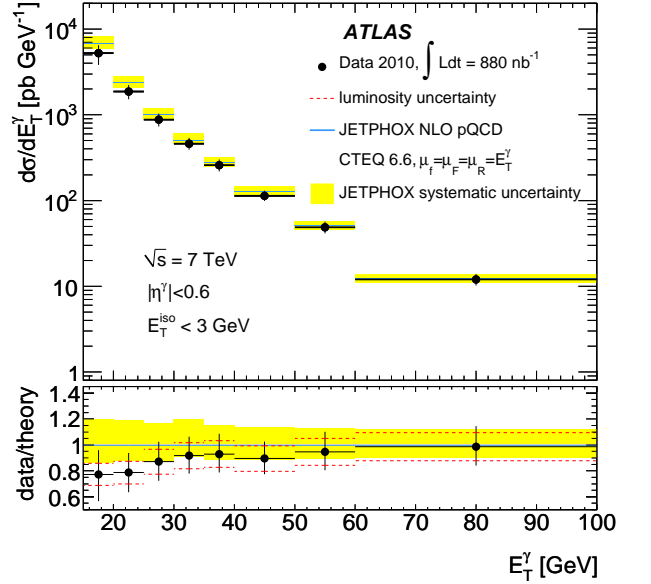}
  \caption{Measured (dots) and expected (full line) inclusive prompt
    photon production cross sections, as a function of the photon
    transverse energies above 15 GeV and in the pseudorapidity range
    $|\eta^\gamma|<0.6$. The bottom panel shows the ratio between the
    measurement and the theoretical prediction.}
  \label{fig:xsec_measured_vs_jetphox_2dsideband_etabin000-060}
\end{figure}

\begin{figure}
  \centering
  \includegraphics[width=1.0\columnwidth]{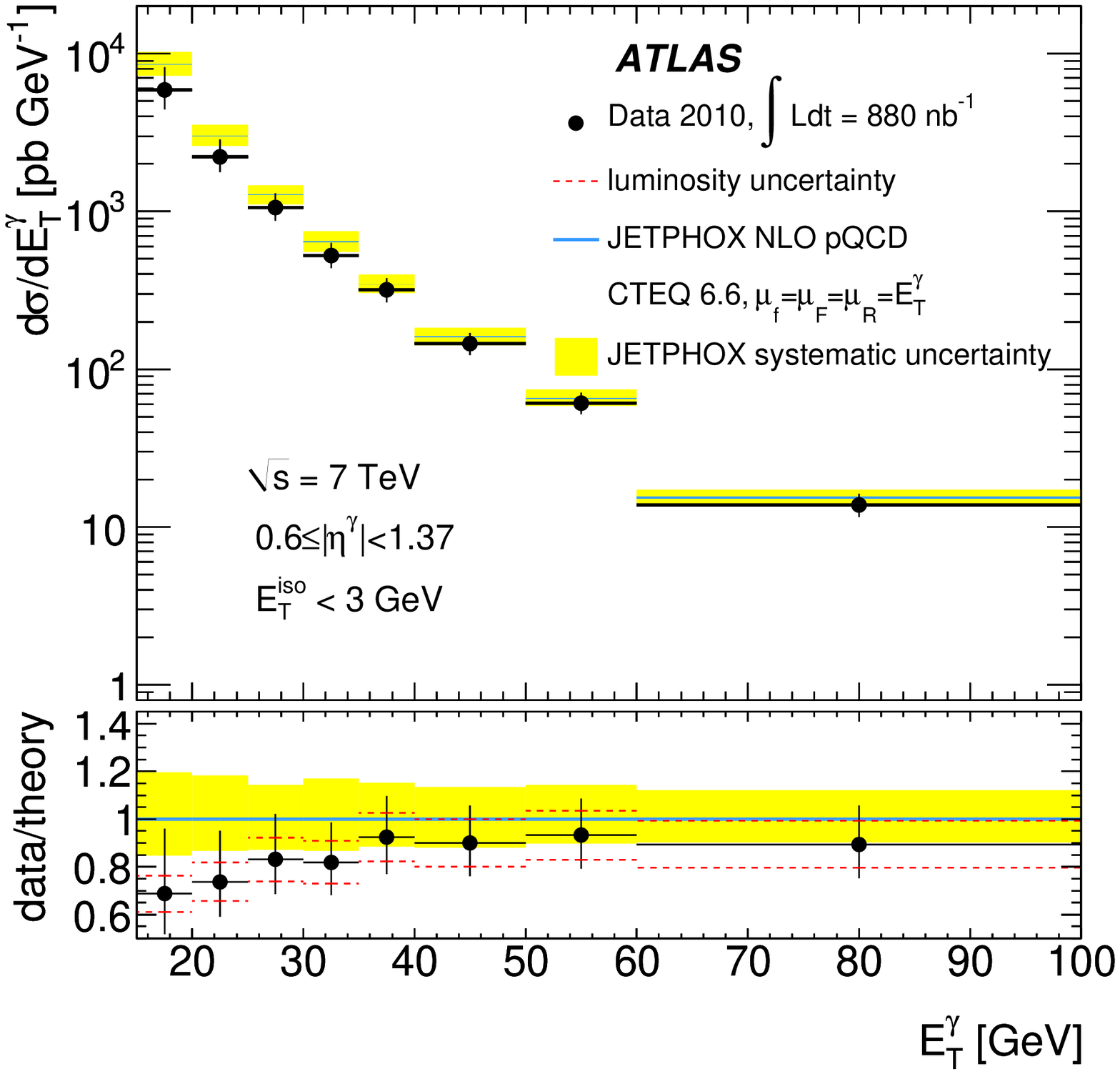}
  \caption{Measured (dots) and expected (full line) inclusive prompt
    photon production cross sections, as a function of the photon
    transverse energies above 15 GeV and in the pseudorapidity range
    $0.6\leq|\eta^\gamma|<1.37$. The bottom panel shows the ratio
    between the measurement and the theoretical prediction.}
  \label{fig:xsec_measured_vs_jetphox_2dsideband_etabin060-137}
\end{figure}

\begin{figure}
  \centering
  \includegraphics[width=1.0\columnwidth]{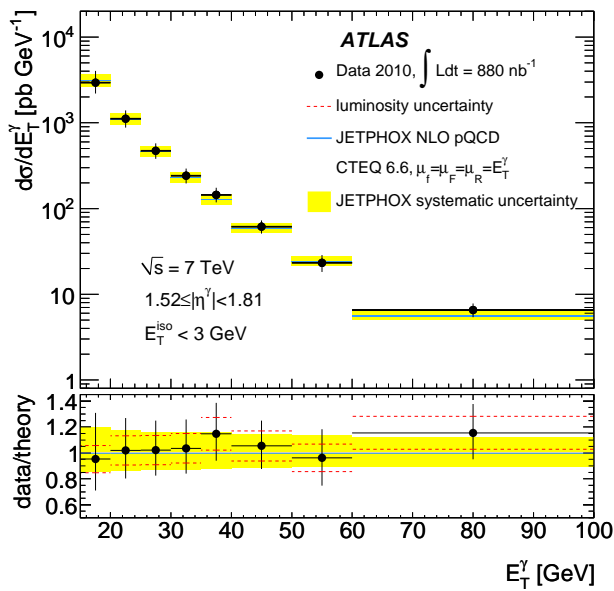}
  \caption{Measured (dots) and expected (full line) inclusive prompt
    photon production cross sections, as a function of the photon
    transverse energies above 15 GeV and in the pseudorapidity range
    $1.52\leq|\eta^\gamma|<1.81$. The bottom panel shows the ratio
    between the measurement and the theoretical prediction.}
  \label{fig:xsec_measured_vs_jetphox_2dsideband_etabin152-181}
\end{figure}

In general, the theoretical predictions agree with the measured cross
sections for $\ET^\gamma>25$~GeV. For lower $\ET$ and in the two
pseudorapidity regions $|\eta^\gamma|<0.6$ and
$0.6\leq|\eta^\gamma|<1.37$, the cross section predicted by {\tt
  JETPHOX} is larger than that measured in data.  Such low transverse
energies at the LHC correspond to extremely small values of 
$x_{\rm T} = 2E_{\rm T}^{\gamma}/\sqrt{s}$, where NLO theoretical predictions are
less accurate.  In such a regime the appropriate values of the
different scales are not clearly defined, and the uncertainties
associated with these scales in the theoretical predictions may not be
well modeled by simple variations of any one scale about the default
value of $E_{\rm T}^\gamma$~\cite{PilonPrivate}.  As the low-$E_{\rm
  T}^{\gamma}$ region is where the fragmentation component has the
most significant contribution to the total cross section, the total
uncertainty associated with the NLO predictions at low $E_{\rm
  T}^{\gamma}$ may be underestimated.

\section{Conclusion}
\label{sec:Conclusion}
The inclusive isolated prompt photon production cross section in $pp$
collisions at a center-of-mass energy $\sqrt{s}=7$ TeV has been
measured using 880~\inb~of $pp$ collision data collected by the ATLAS
detector at the Large Hadron Collider.

The differential cross section has been measured as a function of the
prompt photon transverse energy between 15 and 100 GeV, in the three
pseudorapidity intervals $|\eta^\gamma|<0.6$,
$0.6\leq|\eta^\gamma|<1.37$ and $1.52\leq|\eta^\gamma|<1.81$,
estimating the background from the selected photon sample and
using the photon identification efficiency measurement described
in this paper. The photon identification using the
fine granularity of the calorimeters. A photon isolation criterion is
used, after an {\it in situ} subtraction of the effects of the
underlying event that may also be applied to theoretical predictions.

The observed cross sections rapidly decrease as a function of the
increasing photon transverse energy, spanning almost three orders of
magnitude.  The precision of the measurement is limited by its
systematic uncertainty, which receives important contributions from
the energy scale uncertainty, 
the luminosity, 
the photon identification efficiency, 
and the uncertainty on the residual background
contamination in the selected photon sample.

The NLO pQCD predictions agree with the observed cross sections for
transverse energies greater than 25 GeV, while for transverse energies
below 25 GeV the cross sections predicted by {\tt JETPHOX} are higher
than measured.  However, the precision of this comparison below 25 GeV
is limited by large systematic uncertainties on the measurement and on
the theoretical predictions at such low values of 
$x_{\rm T} = 2E_{\rm T}^{\gamma}/\sqrt{s}$.

The measured prompt photon production cross section is more than a
factor of thirty larger than that measured at the Tevatron, and a
factor of $10^{4}$ larger than for photoproduction at HERA, assuming a
similar kinematic range in transverse energy and pseudorapidity.  This
will allow the extension of the measurement up to energies in the TeV
range after only a few years of data taking at the LHC.

\section{Acknowledgements}

We wish to thank CERN for the efficient commissioning and operation of the LHC during this initial high-energy data-taking period as well as the support staff from our institutions without whom ATLAS could not be operated efficiently.

We acknowledge the support of ANPCyT, Argentina; YerPhI, Armenia; ARC, Australia; BMWF, Austria; ANAS, Azerbaijan; SSTC, Belarus; CNPq and FAPESP, Brazil; NSERC, NRC and CFI, Canada; CERN; CONICYT, Chile; CAS, MOST and NSFC, China; COLCIENCIAS, Colombia; MSMT CR, MPO CR and VSC CR, Czech Republic; DNRF, DNSRC and Lundbeck Foundation, Denmark; ARTEMIS, European Union; IN2P3-CNRS, CEA-DSM/IRFU, France; GNAS, Georgia; BMBF, DFG, HGF, MPG and AvH Foundation, Germany; GSRT, Greece; ISF, MINERVA, GIF, DIP and Benoziyo Center, Israel; INFN, Italy; MEXT and JSPS, Japan; CNRST, Morocco; FOM and NWO, Netherlands; RCN, Norway;  MNiSW, Poland; GRICES and FCT, Portugal;  MERYS (MECTS), Romania;  MES of Russia and ROSATOM, Russian Federation; JINR; MSTD, Serbia; MSSR, Slovakia; ARRS and MVZT, Slovenia; DST/NRF, South Africa; MICINN, Spain; SRC and Wallenberg Foundation, Sweden; SER,  SNSF and Cantons of Bern and Geneva, Switzerland;  NSC, Taiwan; TAEK, Turkey; STFC, the Royal Society and Leverhulme Trust, United Kingdom; DOE and NSF, United States of America.  

The crucial computing support from all WLCG partners is acknowledged gratefully, in particular from CERN and the ATLAS Tier-1 facilities at TRIUMF (Canada), NDGF (Denmark, Norway, Sweden), CC-IN2P3 (France), KIT/GridKA (Germany), INFN-CNAF (Italy), NL-T1 (Netherlands), PIC (Spain), ASGC (Taiwan), RAL (UK) and BNL (USA) and in the Tier-2 facilities worldwide.

\bibliography{InclusiveCrossSection}

\appendix 
\section{Definition of photon identification discriminating variables}

In this Appendix, the quantities used in the selection of photon
candidates, based on the reconstructed energy deposits in the ATLAS
calorimeters, are summarized.

\begin{itemize}

\item {\bf Leakage in the hadronic calorimeter}

  The following discriminating variable is defined, based on the
  energy deposited in the hadronic calorimeter:

\begin{itemize}

\item {\em Normalized hadronic leakage} 
\begin{equation}
\Rhad = \frac{\ET^{\rm had}}{\ET}
\end{equation}
is the total transverse energy $\ET^{\rm had}$ deposited in the hadronic
calorimeter, normalized to the
total transverse energy $\ET$ of the photon candidate.
\end{itemize}
In the $|\eta|$ interval between 0.8 and 1.37 the energy deposited in the
whole hadronic calorimeter is used, while in the other pseudorapidity
intervals only the leakage in the first layer of the hadronic
calorimeter is used.

\item{\bf Variables using the second (``middle'') layer of the
  electromagnetic calorimeter}

The discriminating variables based on the energy deposited in the
second layer of the electromagnetic calorimeter are the following:

\begin{itemize}

\item{\em Middle $\eta$ energy ratio}
\begin{equation}
\Reta = \frac{E_{3\times7}^{S2}}{E_{7\times7}^{S2}}
\end{equation}
is the ratio between the sum $E_{3\times7}^{S2}$ of the energies of
the second layer cells of the electromagnetic calorimeter contained in
a 3$\times$7 rectangle in $\eta\times\phi$ (measured in cell units),
and the sum $E_{7\times7}^{S2}$ of the energies in a 7$\times$7
rectangle, both centered around the cluster seed.

\item{\em Middle $\phi$ energy ratio}
\begin{equation}
\Rphi = \frac{E_{3\times3}^{S2}}{E_{3\times7}^{S2}}
\end{equation}
is defined similarly to $R_{\eta}$. $R_{\phi}$ behaves very
differently for unconverted and converted photons, since the electrons
and positrons generated by the latter bend in different directions in
$\phi$ because of the solenoid magnetic field, producing
larger showers in the $\phi$~direction than the unconverted photons.

\item{\em Middle lateral width}
\begin{equation}
\wetatwo = \sqrt{
\frac{\sum E_i \eta_i^2}{\sum E_i} - \left(\frac{\sum E_i \eta_i}{\sum E_i} \right)^2}
\end{equation}
measures the shower lateral width in the second layer of the
electromagnetic calorimeter, using all cells in a window
$\eta\times\phi = 3 \times 5$ measured in cell units.

\end{itemize}

\item {\bf Variables using the first (``front'') layer of the
  electromagnetic calorimeter}

The discriminating variables based on the energy deposited in the
first layer of the electromagnetic calorimeter are the following:

\begin{itemize}

\item {\em Front side energy ratio}
\begin{equation}
\Fside = \frac{E(\pm 3) - E(\pm 1)}{E(\pm 1)}
\end{equation}
measures the lateral containment of the shower, along the $\eta$
direction.
$E(\pm n)$ is the energy in the $\pm n$ strip cells around the one
with the largest energy. 

\item {\em Front lateral width (3 strips)}
\begin{equation}
\wthree = \sqrt{\frac{\sum E_i (i-i_{\rm max})^2}{\sum E_i}}
\end{equation}
measures the shower width along $\eta$ in the first layer of the
electromagnetic calorimeter, using two strip cells around the maximal
energy deposit.
The index $i$ is the strip identification number, $i_{\rm max}$
identifies the strip cells with the greatest energy, $E_i$ is the
energy deposit in each strip cell.

\item {\em Front lateral width (total)} \\
$\wtot$ measures the shower width along $\eta$ in the first layer of
  the electromagnetic calorimeter using all cells in a window
  $\Delta\eta\times\Delta\phi = 0.0625 \times 0.2$, corresponding 
  approximately to $20\times 2$ strip cells in $\eta\times \phi$, and is
  computed as \wthree. 

\item {\em Front second maximum difference}.
\begin{equation}
\DeltaE = \left[ E_{2^{\rm nd} \rm max}^{S1} - E_{\rm min}^{S1} \right] 
\end{equation}
is the difference between the energy of the strip cell with the
second greatest energy $E_{2^{\rm nd} \rm max}^{S1}$, and the
energy in the strip cell with the least energy found between the
greatest and the second greatest energy $E_{\rm min}^{S1}$ 
($\DeltaE=0$ when there is no second maximum).

\item {\em Front maxima relative ratio}
\begin{equation}
\Eratio = \frac{E_{1^{\rm st}\,\rm max}^{S1}-E_{2^{\rm nd}\,\rm max}^{S1}}
{E_{1^{\rm st}\,\rm max}^{S1}+E_{2^{\rm nd}\,\rm max}^{S1}}
\end{equation}
measures the relative difference between the energy of the strip cell
with the greatest energy $E_{1^{\rm st}\,\rm max}^{S1}$ and the energy
in the strip cell with second greatest energy $E_{2^{\rm nd}\,\rm
max}^{S1}$ (1 when there is no second maximum).
\end{itemize}

\end{itemize}

\label{app:IsEM-DV}

\section{Cross Section Measurements}

Table~\ref{tab:observed_xsec_cteq66_2dsideband_eta000-060},
\ref{tab:observed_xsec_cteq66_2dsideband_eta060-137} and
\ref{tab:observed_xsec_cteq66_2dsideband_eta152-181} list the values
of the measured isolated prompt photon production cross sections, for
the 
$0.00 \leq |\eta^\gamma| < 0.60$, 
$0.60 \leq |\eta^\gamma| < 1.37$ and 
$1.52 \leq |\eta^\gamma| < 1.81$ regions,
respectively.
The various systematic uncertainties originating from the purity
measurement, the photon selection and identification efficiency, the
photon energy scale and the luminosity are shown. The total
uncertainty includes both the statistical and all systematic
uncertainties, except for the uncertainty on the luminosity.

\begin{table*}[hp]
  \centering
  \footnotesize
  \caption{The measured isolated prompt photon production cross section, 
    for $0.00 \leq |\eta^\gamma| < 0.60$.
    The systematic uncertainties originating from the purity measurement,
    the photon selection, the photon energy scale, the unfolding procedure and the luminosity are shown.
    The total uncertainty includes both the statistical and all systematic
    uncertainties, except for the uncertainty on the luminosity.}
  \label{tab:observed_xsec_cteq66_2dsideband_eta000-060}
  \renewcommand{\arraystretch}{1.5}
\begin{tabular}{cllllllllll}
\hline\hline
&                    &\multicolumn{7}{c}{Measured}  &\multicolumn{2}{c}{{\tt JETPHOX}}\\
$\ET^\gamma$         & \mcc{\multirow{2}{*}{$\frac{d\sigma}{dE_{T}^{\gamma}}$}}         & \mcc{stat}         & \mcc{syst}                   & \mcc{syst}                   & \mcc{syst}                  & \mcc{syst}           & \mcc{syst}         & \mcc{total}          &\mcc{\multirow{2}{*}{$\frac{d\sigma}{dE_{T}^{\gamma}}$}}  & \mcc{total}         \\
                     &                                                       &                    & \mcc{(purity)}               & \mcc{(efficiency)}           & \mcc{(en. scale)}              & \mcc{(unfolding)}    & \mcc{(luminosity)}       & \mcc{uncertainty}    &                                               & \mcc{uncertainty}   \\
       $[{\rm GeV}]$ & \mcc{$[{\rm nb/GeV}]$} & \mcc{$[{\rm nb/GeV}]$} & \mcc{$[{\rm nb/GeV}]$}   & \mcc{$[{\rm nb/GeV}]$}       & \mcc{$[{\rm nb/GeV}]$}       & \mcc{$[{\rm nb/GeV}]$}    & \mcc{$[{\rm nb/GeV}]$} & \mcc{$[{\rm nb/GeV}]$}    & \mcc{$[{\rm nb/GeV}]$}      & \mcc{$[{\rm nb/GeV}]$}           \\
\hline
   $[15,  20)$       & 5.24           & $\pm 0.11$   & $^{+0.52}_{-0.88}$     & $\pm 0.81$    & $^{+0.51}_{-0.46}$   & $\pm 0.11$   & $\pm 0.58$   & $^{+1.3}_{-1.4}$       & 6.8     & $^{+1.4}_{-0.9}$                              \\
   $[20,  25)$       & 1.88           & $\pm 0.05$   & $^{+0.18}_{-0.20}$     & $\pm 0.21$    & $^{+0.14}_{-0.14}$   & $\pm 0.04$   & $\pm 0.21$   & $\pm 0.36$           & 2.38     & $^{+0.45}_{-0.30}$                              \\
   $[25,  30)$       & 0.88           & $\pm 0.03$   & $\pm 0.07$           & $\pm 0.08$    & $^{+0.09}_{-0.08}$   & $\pm 0.02$  & $\pm 0.10$   & $^{+0.16}_{-0.15}$            & 1.01     & $^{+0.17}_{-0.13}$                              \\
   $[30,  35)$       & 0.461           & $\pm 0.016$  & $^{+0.029}_{-0.019}$   & $\pm 0.035$   & $^{+0.045}_{-0.046}$ & $\pm 0.009$  & $\pm 0.05$   & $\pm 0.07$          & 0.50     & $^{+0.10}_{-0.04}$                              \\
   $[35,  40)$       & 0.254           & $\pm 0.011$  & $^{+0.017}_{-0.015}$   & $\pm 0.019$   & $^{+0.027}_{-0.025}$ & $\pm 0.005$  & $\pm 0.028$  & $\pm 0.04$          & 0.28     & $^{+0.04}_{-0.03}$                              \\
   $[40,  50)$       & 0.115           & $\pm 0.005$  & $^{+0.008}_{-0.006}$   & $\pm 0.007$   & $^{+0.009}_{-0.009}$ & $\pm 0.0023$ & $\pm 0.013$  & $^{+0.017}_{-0.016}$   & 0.127     & $^{+0.018}_{-0.014}$                              \\
   $[50,  60)$       & 0.050           & $\pm 0.003$  & $^{+0.003}_{-0.002}$   & $\pm 0.003$   & $^{+0.006}_{-0.005}$ & $\pm 0.001$ & $\pm 0.005$  & $^{+0.008}_{-0.007}$   & 0.052     & $^{+0.007}_{-0.006}$                              \\
   $[60, 100)$       & 0.0120           & $\pm 0.0007$ & $^{+0.0007}_{-0.0005}$ & $\pm 0.0006$ & $^{+0.0013}_{-0.0012}$& $\pm 0.0002$ & $\pm 0.0013$ & $^{+0.0019}_{-0.0018}$  & 0.0121     & $^{+0.0014}_{-0.0012}$                              \\
\hline\hline
\end{tabular}
\end{table*}

\begin{table*}[hp]
\centering
\footnotesize
\caption{The measured isolated prompt photon production cross section, for $0.60 \leq |\eta^\gamma| < 1.37$.
  The systematic uncertainties originating from the purity measurement,
  the photon selection, the photon energy scale, the unfolding procedure and the luminosity are shown.
  The total uncertainty includes both the statistical and all systematic
  uncertainties, except for the uncertainty on the luminosity.}
\label{tab:observed_xsec_cteq66_2dsideband_eta060-137}
\renewcommand{\arraystretch}{1.5}
\begin{tabular}{cllllllllll}
\hline\hline
&                    &\multicolumn{7}{c}{Measured}  &\multicolumn{2}{c}{{\tt JETPHOX}}\\
$\ET^\gamma$         & \mcc{\multirow{2}{*}{$\frac{d\sigma}{dE_{T}^{\gamma}}$}}         & \mcc{stat}         & \mcc{syst}                   & \mcc{syst}                   & \mcc{syst}                  & \mcc{syst}           & \mcc{syst}         & \mcc{total}          &\mcc{\multirow{2}{*}{$\frac{d\sigma}{dE_{T}^{\gamma}}$}}  & \mcc{total}         \\
                     &                                                       &                    & \mcc{(purity)}               & \mcc{(efficiency)}           & \mcc{(en. scale)}              & \mcc{(unfolding)}    & \mcc{(luminosity)}       & \mcc{uncertainty}    &                                               & \mcc{uncertainty}   \\
       $[{\rm GeV}]$ & \mcc{$[{\rm nb/GeV}]$} & \mcc{$[{\rm nb/GeV}]$} & \mcc{$[{\rm nb/GeV}]$}   & \mcc{$[{\rm nb/GeV}]$}       & \mcc{$[{\rm nb/GeV}]$}       & \mcc{$[{\rm nb/GeV}]$}    & \mcc{$[{\rm nb/GeV}]$} & \mcc{$[{\rm nb/GeV}]$}    & \mcc{$[{\rm nb/GeV}]$}      & \mcc{$[{\rm nb/GeV}]$}           \\
\hline
   $[15,    20)$       & 5.9           & $\pm 0.2$    & $^{+1.8}_{-0.5}$      & $\pm 1.0$      & $^{+0.6}_{-0.5}$      & $\pm 0.1$    & $\pm 0.6$    & $^{+2.3}_{-1.4}$       & 8.5    & $^{+1.7}_{-1.3}$ \\
   $[20,    25)$       & 2.23           & $\pm 0.07$   & $^{+0.49}_{-0.18}$    & $\pm 0.28$    & $^{+0.16}_{-0.16}$    & $\pm 0.04$   & $\pm 0.24$   & $^{+0.6}_{-0.4}$       & 3.0    & $^{+0.5}_{-0.4}$ \\
   $[25,    30)$       & 1.05           & $\pm 0.03$   & $^{+0.16}_{-0.06}$    & $\pm 0.10$    & $^{+0.10}_{-0.10}$    & $\pm 0.021$  & $\pm 0.12$   & $^{+0.24}_{-0.19}$     & 1.28   & $^{+0.18}_{-0.16}$ \\
   $[30,    35)$       & 0.52           & $\pm 0.02$   & $^{+0.06}_{-0.03}$    & $\pm 0.04$    & $^{+0.05}_{-0.05}$     & $\pm 0.011$  & $\pm 0.06$   & $^{+0.11}_{-0.09}$     & 0.64   & $^{+0.11}_{-0.09}$ \\
   $[35,    40)$       & 0.313           & $\pm 0.014$  & $^{+0.029}_{-0.021}$  & $\pm 0.024$   & $^{+0.035}_{-0.032}$   & $\pm 0.006$  & $\pm 0.034$  & $^{+0.06}_{-0.05}$     & 0.344  & $^{+0.052}_{-0.039}$ \\
   $[40,    50)$       & 0.146           & $\pm 0.006$  & $^{+0.014}_{-0.011}$  & $\pm 0.009$   & $^{+0.013}_{-0.013}$   & $\pm 0.003$  & $\pm 0.016$  & $^{+0.025}_{-0.022}$   & 0.161  & $^{+0.022}_{-0.019}$ \\
   $[50,    60)$       & 0.062           & $\pm 0.004$  & $^{+0.005}_{-0.004}$  & $\pm 0.003$   & $^{+0.006}_{-0.006}$   & $\pm 0.001$  & $\pm 0.007$  & $^{+0.010}_{-0.009}$   & 0.065  & $^{+0.009}_{-0.007}$ \\
   $[60,   100)$       & 0.0138           & $\pm 0.0008$ & $^{+0.0013}_{-0.0009}$ & $\pm 0.0007$ & $^{+0.0016}_{-0.0014}$ & $\pm 0.0003$ & $\pm 0.0015$ & $^{+0.0025}_{-0.0022}$ & 0.0154 & $^{+0.0019}_{-0.0015}$ \\
\hline\hline
\end{tabular}
\end{table*}
\begin{table*}[tbp]
\centering
\footnotesize
\caption{The measured isolated prompt photon production cross section, for $1.52 \leq |\eta^\gamma| < 1.81$.
The systematic uncertainties originating from the purity measurement,
the photon selection, the photon energy scale, the unfolding procedure and the luminosity are shown.
The total uncertainty includes both the statistical and all systematic
uncertainties, except for the uncertainty on the luminosity.}
\label{tab:observed_xsec_cteq66_2dsideband_eta152-181}
\renewcommand{\arraystretch}{1.5}
\begin{tabular}{cllllllllll}
\hline\hline
&                    &\multicolumn{7}{c}{Measured}  &\multicolumn{2}{c}{{\tt JETPHOX}}\\
$\ET^\gamma$         & \mcc{\multirow{2}{*}{$\frac{d\sigma}{dE_{T}^{\gamma}}$}}         & \mcc{stat}         & \mcc{syst}                   & \mcc{syst}                   & \mcc{syst}                  & \mcc{syst}           & \mcc{syst}         & \mcc{total}          &\mcc{\multirow{2}{*}{$\frac{d\sigma}{dE_{T}^{\gamma}}$}}  & \mcc{total}         \\
                     &                                                       &                    & \mcc{(purity)}               & \mcc{(efficiency)}           & \mcc{(en. scale)}              & \mcc{(unfolding)}    & \mcc{(luminosity)}       & \mcc{uncertainty}    &                                               & \mcc{uncertainty}   \\
       $[{\rm GeV}]$ & \mcc{$[{\rm nb/GeV}]$} & \mcc{$[{\rm nb/GeV}]$} & \mcc{$[{\rm nb/GeV}]$}   & \mcc{$[{\rm nb/GeV}]$}       & \mcc{$[{\rm nb/GeV}]$}       & \mcc{$[{\rm nb/GeV}]$}    & \mcc{$[{\rm nb/GeV}]$} & \mcc{$[{\rm nb/GeV}]$}    & \mcc{$[{\rm nb/GeV}]$}      & \mcc{$[{\rm nb/GeV}]$}           \\
\hline
   $[15,    20)$       & 2.9              & $\pm 0.1$    & $^{+0.8}_{-0.3}$      & $\pm 0.5$      & $^{+0.3}_{-0.3}$      & $\pm 0.1$   & $\pm 0.3$    & $^{+1.1}_{-0.7}$      & 3.1    & $^{+0.6}_{-0.5}$\\
   $[20,    25)$       & 1.12             & $\pm 0.04$   & $^{+0.15}_{-0.08}$    & $\pm 0.16$     & $^{+0.08}_{-0.08}$    & $\pm 0.02$  & $\pm 0.12$   & $^{+0.27}_{-0.24}$    & 1.10   & $^{+0.20}_{-0.15}$\\
   $[25,    30)$       & 0.47             & $\pm 0.02$   & $^{+0.06}_{-0.04}$    & $\pm 0.05$    & $^{+0.05}_{-0.04}$    & $\pm 0.01$  & $\pm 0.05$   & $^{+0.11}_{-0.09}$    & 0.46   & $^{+0.07}_{-0.06}$\\
   $[30,    35)$       & 0.240            & $\pm 0.013$  & $^{+0.028}_{-0.016}$  & $\pm 0.023$   & $^{+0.025}_{-0.026}$  & $\pm 0.005$  & $\pm 0.026$  & $^{+0.052}_{-0.045}$   & 0.233   & $^{+0.037}_{-0.030}$\\
   $[35,    40)$       & 0.142            & $\pm 0.009$  & $^{+0.018}_{-0.010}$  & $\pm 0.012$   & $^{+0.014}_{-0.013}$  & $\pm 0.0032$ & $\pm 0.016$  & $^{+0.030}_{-0.026}$  & 0.126  & $^{+0.020}_{-0.015}$\\
   $[40,    50)$       & 0.062            & $\pm 0.004$  & $^{+0.005}_{-0.004}$  & $\pm 0.005$   & $^{+0.006}_{-0.006}$  & $\pm 0.0013$ & $\pm 0.007$  & $^{+0.011}_{-0.010}$   & 0.058  & $^{+0.008}_{-0.007}$\\
   $[50,    60)$       & 0.0237           & $\pm 0.0025$ & $^{+0.0026}_{-0.0028}$ & $\pm 0.0019$ & $^{+0.0024}_{-0.0022}$& $\pm 0.0005$ & $\pm 0.003$  & $\pm 0.005$   & 0.0243 & $^{+0.0033}_{-0.0027}$\\
   $[60,   100)$       & 0.0066           & $\pm 0.0005$ & $^{+0.0005}_{-0.0003}$ & $\pm 0.0005$ & $^{+0.0008}_{-0.0007}$& $\pm 0.0002$ & $\pm 0.0007$ & $^{+0.0013}_{-0.0012}$ & 0.0057 & $^{+0.0007}_{-0.0006}$\\
\hline\hline
\end{tabular}
\end{table*}

\label{app:CS-tables}

\clearpage
\onecolumngrid
\section{The ATLAS Collaboration}
\label{app:ATLASColl}
\begin{center}
\begin{flushleft}
{\Large The ATLAS Collaboration}

\bigskip

G.~Aad$^{\rm 48}$,
B.~Abbott$^{\rm 111}$,
J.~Abdallah$^{\rm 11}$,
A.A.~Abdelalim$^{\rm 49}$,
A.~Abdesselam$^{\rm 118}$,
O.~Abdinov$^{\rm 10}$,
B.~Abi$^{\rm 112}$,
M.~Abolins$^{\rm 88}$,
H.~Abramowicz$^{\rm 153}$,
H.~Abreu$^{\rm 115}$
% ...
\bigskip

% ...
$^{10}$ Institute of Physics, Azerbaijan Academy of Sciences, H. Javid Avenue 33, AZ 143 Baku, Azerbaijan\\
$^{11}$ Institut de F\'isica d'Altes Energies, IFAE, Edifici Cn, Universitat Aut\`onoma  de Barcelona,  ES - 08193 Bellaterra (Barcelona), Spain\\
% ...
$^{48}$ Albert-Ludwigs-Universit\"{a}t, Fakult\"{a}t f\"{u}r Mathematik und Physik, Hermann-Herder Str. 3, D - 79104 Freiburg i.Br., Germany\\
$^{49}$ Universit\'e de Gen\`eve, Section de Physique, 24 rue Ernest Ansermet, CH - 1211 Geneve 4, Switzerland\\
% ...
$^{88}$ Michigan State University, Department of Physics and Astronomy, High Energy Physics Group, East Lansing, MI 48824-2320, United States of America\\
% ...
$^{111}$ University of Oklahoma, Homer L. Dodge Department of Physics and Astronomy, 440 West Brooks, Room 100, Norman, OK 73019-0225, United States of America\\
$^{112}$ Oklahoma State University, Department of Physics, 145 Physical Sciences Building, Stillwater, OK 74078-3072, United States of America\\
% ...
$^{115}$ LAL, Univ. Paris-Sud, IN2P3/CNRS, Orsay, France\\
% ...
$^{118}$ Oxford University, Department of Physics, Denys Wilkinson Building, Keble Road, Oxford OX1 3RH, United Kingdom\\
% ...
$^{153}$ Tel Aviv University, Raymond and Beverly Sackler School of Physics and Astronomy, Ramat Aviv, IL - Tel Aviv 69978, Israel\\
% ...
\end{flushleft}


\end{center}

\end{document}